\documentclass[fleqn,usenatbib]{mnras}

\usepackage{newtxtext,newtxmath}
\usepackage[T1]{fontenc}

\DeclareRobustCommand{\VAN}[3]{#2}
\let\VANthebibliography\thebibliography
\def\thebibliography{\DeclareRobustCommand{\VAN}[3]{##3}\VANthebibliography}

\usepackage{graphicx}	% Including figure files
\usepackage[normalem]{ulem}

\usepackage{tabularx}
\usepackage{multirow}
\usepackage{xcolor}

\usepackage{graphicx}	% Including figure files
\usepackage{amsmath}	% Advanced maths commands
\newcommand{\Msun}{M$_{\odot}$}

\newcommand{\DP}{$\Delta P/P_\mathrm{DM}$}
\newcommand{\BF}{$\overline{f}_\mathrm{bar}$}
\newcommand{\NHj}{$N^{j}_\mathrm{halo}$}
\newcommand{\NH}{$N_\mathrm{halo}$}
\newcommand{\vDBF}{$\overline{f}_\mathrm{bar}(M_{\rm{halo}}>13.5)$}
\newcommand{\BinBF}{$\overline{f}_\mathrm{bar}(M^{\rm j}_{\rm{halo}})$}
\newcommand{\PhPd}{$P_{\rm{hydro}}$/$P_{\rm{DM}}$}

\newcommand{\Om}{$\Omega_{\rm{m}}$}
\title[Matter clustering in CAMELS]{Predicting the impact of feedback on matter clustering with machine learning in CAMELS}

\author[Delgado et al.]{Ana Maria Delgado,$^{1}$\thanks{E-mail: ana\_maria.delgado@cfa.harvard.edu (AMD)}
Daniel Angl\'es-Alc\'azar,$^{2,3}$
Leander Thiele,$^{4}$
Shivam Pandey,$^{7,8}$
Kai Lehman,$^{9,10}$
\newauthor%
Rachel S. Somerville,$^{3}$
Michelle Ntampaka,$^{5,6}$
Shy Genel,$^{3}$
Francisco Villaescusa-Navarro$^{3}$
Lars Hernquist,$^{1}$
\\
$^{1}$Center for Astrophysics, Harvard and Smithsonian, 60 Garden Street, Cambridge, MA, 02138 USA\\
$^{2}$Department of Physics, University of Connecticut, 196 Auditorium Road, U-3046, Storrs, CT, 06269, USA \\
$^{3}$Center for Computational Astrophysics, Flatiron Institute, 162 5th Avenue, New York, NY, 10010, USA\\
$^{4}$Department of Physics, Princeton University, Jadwin Hall, Princeton, NJ, 08544, USA\\
$^{5}$Data Science Mission Office, Space Telescope Science Institute, Baltimore, MD, 21218, USA\\
$^{6}$Department of Physics and Astronomy, Johns Hopkins University, Baltimore, MD, 21218, USA\\
$^{7}$Department of Physics, Colombia University, New York, NY, USA\\
$^{8}$Department of Physics and Astronomy, University of Pennsylvania, Philadelphia, PA, 19104, USA\\
$^{9}$Institute for Astronomy, University of Hawai'i, 2680 Woodlawn drive, Honolulu, HI, 96822, USA\\
$^{10}$Universit\"{a}ts-Sternwarter M\"{u}nchen, Fakult\"{a}t f\"{u}r Physik, Ludwig-Maximilians-Universit\"{a}t, Scheinerstr. 1, 81679, M\"{u}nchen, Germany
\\
}

\date{Accepted XXX. Received YYY; in original form ZZZ}

\pubyear{2021}

\begin{document}
\label{firstpage}
\pagerange{\pageref{firstpage}--\pageref{lastpage}}
\maketitle

\begin{abstract}

Extracting information from the total matter power spectrum with the precision needed for upcoming cosmological surveys requires unraveling the complex effects of galaxy formation processes on the distribution of matter. We investigate the impact of baryonic physics on matter clustering at $z=0$ using a library of power spectra from the Cosmology and Astrophysics with MachinE Learning Simulations (CAMELS) project, containing thousands of $(25\,h^{-1}{\rm Mpc})^3$ volume realizations with varying cosmology, initial random field, stellar and AGN feedback strength and sub-grid model implementation methods. 
We show that baryonic physics affects matter clustering on scales $k \gtrsim 0.4\,h\,\mathrm{Mpc}^{-1}$ and the magnitude of this effect is dependent on the details of the galaxy formation implementation and variations of cosmological and astrophysical parameters.
Increasing AGN feedback strength decreases halo baryon fractions and yields stronger suppression of power relative to N-body simulations, while stronger stellar feedback often results in weaker effects by suppressing black hole growth and therefore the impact of AGN feedback.
We find a broad correlation between mean baryon fraction of massive halos ($M_{\rm 200c} > 10^{13.5}$\,\Msun) and suppression of matter clustering but with significant scatter compared to previous work owing to wider exploration of feedback parameters and cosmic variance effects.
We show that a random forest regressor trained on the baryon content and abundance of halos across the full mass range $10^{10} \leq M_\mathrm{halo}/$\Msun$< 10^{15}$ can predict the effect of galaxy formation on the matter power spectrum on scales $k = 1.0$--20.0\,$h\,\mathrm{Mpc}^{-1}$.

\end{abstract}


%%%%%%%%%%%%%%%%%%%%%%%%%%%%%%%%%%%%%%%%%%%%%%%%%%

%%%%%%%%%%%%%%%%% BODY OF PAPER %%%%%%%%%%%%%%%%%%

\section{Introduction}
\label{Intro}

The field of cosmology has many exciting endeavors to look forward to within the next decade. With the arrival of enormous photometric and spectroscopic galaxy redshift survey missions such as DESI \citep{2016arXiv161100036D}, the Nancy Roman Space Telescope \citep{2015arXiv150303757S}, Euclid \citep{2011arXiv1110.3193L} and the Vera Rubin Observatory \citep{2009arXiv0912.0201L}, the community will have the opportunity to tackle many ambitious goals, such as mapping the distribution of matter and the large scale structure of the Universe, measuring cosmological parameters to percent-level precision, and constraining the sum of neutrino masses. An important step in fully realizing the statistical power of these upcoming surveys is to model the matter power spectrum and other summary statistics to $\sim$1\% precision down to scales as small as $k=10\,h$\,Mpc$^{-1}$ \citep{2005APh....23..369H, 2009arXiv0912.0914L, 2012JCAP...04..034H}.  
However, previous studies have shown that complex galaxy formation processes involving feedback from massive stars and active galactic nuclei (AGN) can suppress power relative to dark matter-only simulations out to large scales \citep{2011MNRAS.415.3649V, 2020MNRAS.491.2424V, 2018MNRAS.480.3962C, 2023arXiv230711832G}. Galactic winds driven by supernovae and AGN-driven outflows can eject a large amount of material from the center of galaxies out to large distances \citep{2017MNRAS.470.4698A,Borrow2020,Hafen2020_CGMfates,Wright2020,Ayromlou2023,Mitchell2022,Sorini2022} and the resulting suppression of power by feedback creates significant biases when attempting to constrain cosmological parameters \citep{2011MNRAS.417.2020S, PhysRevD.87.043509, 2019OJAp....2E...4C}.

Several approaches to addressing the suppression of matter clustering caused by baryonic physics have been devised. 
Cosmological hydrodynamic simulations provide the most direct method to understand the impact of baryonic effects on the distribution and clustering of matter \citep{2014Natur.509..177V, 2016MNRAS.461L..11H, 2015MNRAS.453..469T, 2018MNRAS.475..676S, 2019OJAp....2E...4C}. Modern cosmological large-volume simulations such as Horizon-AGN \citep{2014MNRAS.444.1453D}, Eagle \citep{Schaye2015}, IllustrisTNG \citep{2018MNRAS.475..648P, 2018MNRAS.475..676S, 2018MNRAS.475..624N, 2018MNRAS.477.1206N, 2018MNRAS.480.5113M}, and SIMBA \citep{2019MNRAS.486.2827D} produce galaxies that broadly match observations in properties such as the stellar mass function and the bimodality in galaxy colors. Comparing the power spectrum of hydrodynamic simulations with those of their phase-matched, collisonless N-body, dark matter-only simulations allows us to measure how baryonic feedback suppresses the clustering of matter. However, many key feedback processes remain poorly understood and most current models require extensive tuning of free parameters to match observations, limiting their predictive power \citep{SomervilleDave2015}. Higher resolution cosmological ``zoom-in'' simulations can reduce subgrid model uncertainties \citep[e.g.,][]{Agertz2016,Hopkins2018_FIRE2methods,Angles-Alcazar2021}, but at the expense of modeling volumes that are too small for many cosmological applications.

More flexible approaches to address the impact of baryonic physics using analytic models include: modifying the ``halo model'' \citep{2000MNRAS.318..203S, 2013MNRAS.434..148S, 2014JCAP...04..028F, 2015MNRAS.454.1958M} using observational constraints and simulation results as the basis for  parameterizing the transfer of power produced by the presence of baryons \citep{2014MNRAS.445.3382M, 2015JCAP...12..049S}, and mitigating the presence of baryons altogether by marginalizing over the parameters of effective models \citep{2011MNRAS.417.2020S} or over the principle components in linear combinations of observables that are most strongly affected by baryonic effects \citep{2015MNRAS.454.2451E, 2016MNRAS.459..971K}. However, the success of these techniques rely heavily on the flexibility of the models to capture the true underlying distribution of matter \citep{2017MNRAS.465.2936M} and they are limited by assumptions about halo bias relative to the linear density field, smooth halo profiles neglecting substructure, and uncertainties in the spatial and redshift dependence of baryonic effects \citep{2019OJAp....2E...4C}.
Alternatively, power spectra produced by a large number of cosmological simulations with varying cosmologies and feedback parameters can be used to inform semi-analytic models attempting to mitigate the effects of baryons, characterize the theoretical uncertainties in galaxy formation, and marginalize over feedback effects.

\citet{2011MNRAS.415.3649V} employed a suite of 50 cosmological hydrodynamic simulations from the OWLS project \citep{Schaye2010} to study the effects of different baryonic processes on the matter power spectrum over a range of scales. More recently, \cite{2020MNRAS.491.2424V} (henceforth vDMS) included additional simulations from the cosmo-OWLS \citep{LeBrun2014} and BAHAMAS \citep{2017MNRAS.465.2936M} projects to produce a library of 92 matter power spectra from simulations with varying subgrid models and feedback strengths.
Relating the effects of galaxy formation physics to the suppression of power, vDMS proposed that it is possible to predict the fractional impact of baryons on the clustering of matter, $P_\mathrm{hydro}/P_\mathrm{DM}$, given only the mean baryon fraction of massive halos ($M_{\rm halo} \sim 10^{14}$\,\Msun), where $P_\mathrm{hydro}$ and $P_\mathrm{DM}$ are the matter power spectra from hydrodynamic simulations and their corresponding N-body simulations, respectively. 
Importantly, the empirical vDMS relation between baryon fraction and power suppression is satisfied by a variety of simulations with different galaxy formation implementations, including the Horizon-AGN, EAGLE, and IllustrisTNG simulations, which opens the possibility to accurately correct dark matter only power spectra based on observational constraints on gas fractions in massive halos. 
However, this relation is valid only on large scales, $k \leq 1\,h\,{\rm Mpc}^{-1}$, and the still limited number of different feedback implementations and cosmologies represented in the vDMS library of matter power spectra may not be representative of a broader range of plausible galaxy formation models.

In this work, we use 2,000+ cosmological hydrodynamic simulations and their corresponding collisionless (N-body) simulations from the Cosmology and Astrophysics with MachinE Learning Simulations (CAMELS\footnote{\url{https://www.camel-simulations.org}}) project \citep{camels_presentation} to examine the impact of baryonic physics on matter clustering using the largest library of power spectra available including variations of cosmological and feedback parameters.  
In recent related work using CAMELS, \cite{2022JCAP...04..046N} trained a neural network on thousands of electron density auto-power spectra from large scales down to $k = 10\,h\,\mathrm{Mpc}^{-1}$, breaking the baryon-cosmology degeneracy and providing tight constraints on the total matter density $\Omega_{\rm m}$ and the mean baryon fraction in intermediate-mass halos while marginalizing over uncertainties in galaxy formation physics implementations.
Here, we significantly expand upon the work of vDMS and investigate how supernova and AGN feedback affect the mean baryon fraction across a range of halo masses ($10^{10} \leq M_\mathrm{halo}/$\Msun$ < 10^{15}$) and the resulting impact on the matter power spectrum. Furthermore, we take advantage of the design of CAMELS for machine learning and train a random forest regressor to predict the relative difference between the matter clustering in hydrodynamical and N-body simulations on scales $k = 1.0$--20\,$h\,\mathrm{Mpc}^{-1}$ given the mean baryon fraction of halos across a broad range of halo masses. We thus demonstrate that we are able to extract valuable information from lower mass halos and predict the suppression of power all the way to the highly non-linear regime. The work presented here is complementary to \cite{2023arXiv230102186P}, which show that information about the impact of baryonic effects on the matter power spectrum can be extracted using the tSZ signals from low-mass halos, and include related results utilizing the suite of CAMELS produced with the Astrid simulation.

The layout of this paper is as follows: In section \ref{Methods} we describe the simulations and halo selection, define our variables, and describe our machine learning methods. In sections \ref{Results: feedback} and \ref{Results:ML} we present our results. Finally, in section \ref{summary} we provide a summary and discussion of our work.

%%%%%%%%%%%%%%%%%%%%%%%%%%%%%%%%%%%%%%%%%%%%%%%%%%
\section{Methods}
\label{Methods}
\subsection{Simulations: CAMELS}
\label{Methods:CAMELS}
The CAMELS project \citep{camels_presentation} contains thousands of state-of-the-art (magneto-)hydrodynamic simulations and their corresponding N-body simulations.  In this work, we focus on the simulation suites produced with the IllustrisTNG \citep{2018MNRAS.475..648P, 2018MNRAS.475..676S, 2018MNRAS.475..624N, 2018MNRAS.477.1206N, 2018MNRAS.480.5113M} and SIMBA \citep{2019MNRAS.486.2827D} galaxy formation models that are part of the CAMELS public data release\footnote{\url{https://camels.readthedocs.io}} \citep{Villaescusa-Navarro2022_CAMELSpublic}, providing a total of $>$4000 realizations with parameter variations, with $>$1000 hydrodynamical and $>$1000 N-body simulations for each of the two independent feedback model implementations. Each simulation is a periodic box of length $L_{\mathrm{box}}=25\,h^{-1}\mathrm{Mpc}$ containing $256^3$ resolution elements with mass resolution of $6.49 \times 10^7\ (\Omega_{\rm m}-\Omega_{\rm b})/0.251\,h^{-1}\rm{M}_{\odot}$ for dark matter and $1.27 \times 10^7\,h^{-1}\rm{M}_{\odot}$ for baryons. This is the same resolution as the original SIMBA simulation and similar to that of the original TNG300-1 simulation of IllustrisTNG. 

The initial conditions of CAMELS simulations were generated at $z=127$ using second order Lagrangian perturbation theory assuming that the initial power spectra of dark matter and gas are the same and equal to that of total matter. 
Each of the CAMELS simulations contains 34 snapshots from redshifts $z=6$ down to $z=0$; in this work we focus on $z=0$. In addition to the initial random phases, each simulation is specified by two cosmological parameters and four astrophysical (feedback) parameters which are varied across the individual realizations. In the case of cosmological parameters, we vary:
\begin{itemize}
    \item \textbf{$\Omega_{\rm m}$}: the fraction of the Universe made up of ordinary and dark matter varies in the range $\Omega_{\rm m} \in [0.1,0.5]$ while keeping $\Omega_{\rm b} = 0.049$ constant.\\
    
    \item \textbf{$\sigma_8$}: the variance of the spatial fluctuations of total matter on 8\,Mpc\,$h^{-1}$ scales is varied in the range $\sigma_8 \in [0.6,1.0]$.
\end{itemize}

In the case of astrophysical parameters, the fiducial values are defined by the stellar and AGN feedback models of the corresponding original IllustrisTNG and SIMBA simulations. The fiducial astrophysical parameters are assigned a value $A = 1.0$ and then varied across realizations by multiplying by an amplitude factor $A$ in order to increase/decrease the amount of feedback. However, we emphasize that the stellar and AGN feedback prescriptions differ substantially between IllustrisTNG and SIMBA and the corresponding parameter variations in CAMELS have a different definition in each model, which we briefly describe below.

\subsubsection{IllustrisTNG}
The IllustrisTNG model \citep[also referred to as ``TNG'';][]{2018MNRAS.475..648P, 2018MNRAS.475..676S, 2018MNRAS.475..624N, 2018MNRAS.477.1206N, 2018MNRAS.480.5113M} is implemented in the AREPO hydrodynamics code \citep{Springel_2010, Weinberger2020}, which utilizes a hybrid tree/particle-mesh scheme to solve for gravitational interactions and an unstructured, moving mesh to solve the equations of hydrodynamics. Compared to the galaxy formation model of its predecessor \textit{Illustris} \citep{2014MNRAS.444.1518V, 2014Natur.509..177V, 2014MNRAS.445..175G}, the galaxy formation model in IllustrisTNG has updated implementations of AGN feedback \citep{2017MNRAS.465.3291W} and galactic winds \citep{2018MNRAS.473.4077P}, and incorporates magnetic fields \citep{Pakmor_2014}.

The stellar feedback parameter variations in the CAMELS-TNG simulations introduce \textbf{$A_{\rm{SN1}}$} to control the total energy injection rate in galactic winds per unit star formation ($A_{\rm{SN1}} \in [0.25,4.0]$) and \textbf{$A_{\rm{SN2}}$} to vary the galactic wind speed ($A_{\rm{SN2}} \in [0.5,2.0]$). The AGN feedback parameter variations pertain to the low-accretion, kinetic-mode black hole feedback, where \textbf{$A_{\rm{AGN1}}$} varies the feedback energy per unit black hole accretion rate ($A_{\rm{AGN1}} \in [0.25,4.0]$) and \textbf{$A_{\rm{AGN2}}$} varies the burstiness and effective ejection speed ($A_{\rm{AGN2}} \in [0.5,2.0]$).

\subsubsection{SIMBA}
The SIMBA galaxy formation model \citep{2019MNRAS.486.2827D} is implemented in the GIZMO meshless finite mass hydrodynamics code \citep{2015MNRAS.450...53H, 2017arXiv171201294H}. Relative to its predecessor MUFASA \citep{2016MNRAS.462.3265D}, SIMBA includes a black hole model based on gravitational torque accretion and two-mode kinetic feedback \citep{Angles-Alcazar2017_BHfeedback}, galactic winds with mass-loading and velocity scalings derived from the FIRE zoom-in simulations \citep{Muratov2015,2017MNRAS.470.4698A}, and a model for the creation and destruction of dust \citep{Li2019_SimbaDust}. 

The stellar feedback parameter variations in the CAMELS-SIMBA simulations introduce \textbf{$A_{\rm{SN1}}$} to control the mass loading factor of galactic winds and \textbf{$A_{\rm{SN2}}$} to control the wind speed. The AGN feedback parameter variations introduce \textbf{$A_{\rm{AGN1}}$} to change the total momentum flux of either quasar-mode winds or radio-mode jets, while \textbf{$A_{\rm{AGN2}}$} controls the maximum velocity of gas ejected by jets. These parameters are varied over the same range as in IllustrisTNG, with $A_{\rm{SN1}} = A_{\rm{SN2}} = A_{\rm{AGN1}} = A_{\rm{AGN2}} = 1$ corresponding to the fiducial model. 
As described in \citet{camels_presentation}, the range of feedback parameters explored in CAMELS was chosen to roughly produce factor-of-two variations of injected feedback energy relative to the fiducial models, as a compromise between investigating a wide range of feedback effects while still considering physically plausible models. We also stress that despite using the same range of parameter variations in IllustrisTNG and SIMBA, the resulting effects are model dependent (as shown below and previous works) and reflect their specific implementation.

\subsubsection{Simulation sets in CAMELS}
We take advantage of the following simulation sets for each of the IllustrisTNG and SIMBA suites in CAMELS:
\begin{itemize}
    \item Latin Hypercube (``{\bf LH}'') set: 1,000 realizations each containing different initial conditions and different values of the six aforementioned parameters. The {\bf LH} set is the main training set in this work.\\
    
    \item 1 Parameter (``{\bf 1P}'') set: 66 realizations using the same initial conditions and further divided into six subsets of 11 realizations where only the value of one parameter is varied while the other five parameters are held constant. In this work we make use of the {\bf 1P} sets to study how a single cosmological or feedback parameter can affect halo baryon fractions and the suppression of the matter power spectrum.\\
    
    \item Cosmic Variance (``{\bf CV}'') set: 27 realizations with different initial conditions while the fiducial values of all six parameters are held constant. The {\bf CV} set is used to evaluate the impact of cosmic variance on any of the quantities that we measure from the simulations. 
\end{itemize}

We refer the reader to \cite{camels_presentation} for further details about CAMELS, the parameter variations, and the simulation sets available.

\subsection{Halo Selection}
\label{Methods: halo selection}
We identify halos in CAMELS using the AMIGA Halo Finder \citep[AHF;][]{2011ascl.soft02009K}. AHF uses an adaptive mesh to locate halo centers, calculate the gravitational potential of the halo and iteratively remove unbound particles (particles whose velocities are greater than the escape velocity at a given radius) from within the boundary of the halo. We refer the reader to (\cite{2011ascl.soft02009K}) for a full description and implementation of AHF. We select halos with masses $M_{\mathrm{halo}} \geq 10^{10}$ \Msun~using a virial radius definition of 200c (i.e., 200 times the critical density of the Universe).

\subsection{Matter power spectra and halo baryon fractions}
\label{Methods: variables}

We use a library of $>$4,000 total matter power spectra from CAMELS \citep{2023ApJS..265...54V}. For each simulation, the matter power spectrum is computed by assigning particle masses (dark matter, gas, stars, and black holes) to a regular grid with $512^3$ voxels using a cloud-in-cell (CIC) assignment scheme. The grid is then Fourier transformed and the power spectrum is computed by averaging over $k$-bins with an equal width to the fundamental frequency, $k_{\rm{F}}=2\pi /L$, where $L=25h^{-1}$Mpc. We then compute the relative difference between the total matter power spectrum of hydrodynamical and phase-matched N-body simulations, which we refer to as the ``suppression of matter power spectrum'' and define as:
\begin{equation}
        \frac{\Delta P}{P_\mathrm{DM}} = \frac{P_\mathrm{hydro}-P_\mathrm{DM}}{P_\mathrm{DM}},
\end{equation}
where $P_\mathrm{DM}$ is the matter power spectrum of the N-body simulation and $P_\mathrm{hydro}$ is that of its corresponding hydrodynamical simulation.

We compute the baryon fraction of a given halo as:
\begin{equation}\label{equ:f_bar}
        f_\mathrm{bar}=\frac{M_\mathrm{star} + M_\mathrm{gas}}{M_\mathrm{halo}},
\end{equation}
where $M_{\rm{star}}$ and $M_{\rm{gas}}$ are the total stellar mass and gas mass of the halo and $M_{\rm{halo}}$ is the virial mass of the halo corresponding to $R_{\rm{200c}}$. We further calculate the mean baryon fraction within a given halo mass range in each simulation as:
\begin{equation}\label{eq:fbar}
        \overline{f}_{\rm{bar}}= \frac{1}{n}\sum_{i=1}^{n} f_{\rm{bar_i}} / \frac{\Omega_{\rm b}}{\Omega_{\rm m}},
\end{equation}
where $f_{\rm{bar}}$ is defined in Equation \ref{equ:f_bar}, subscript $i$ is the $i^{\rm th}$ halo and $n$ the total number of halos in a given mass range, and following vDMS we normalize by $\Omega_{\rm b}/ \Omega_{\rm m}$ in order to account for the differences in cosmology for different simulations.

\subsection{Machine Learning}
\label{Methods :ML}
A supervised machine learning algorithm trains a model by providing a subset of data, referred to as the training set, including input variables (henceforth called ``features'') and output variables (henceforth called ``target''). The goal is for the algorithm to use the training set to learn the relation between the features and the target. The trained model is then used to predict the target for a different subset of features referred to as the test set.

In this work we use the random forest regressor algorithm from the publicly available package \texttt{Scikit-Learn} \citep{Scikit}. A random forest (RF) is an ensemble machine learning method that can be used for both classification and regression problems. 
The algorithm works by constructing a ``forest'' from a user specified number of decision trees and using the mean of the predictions from those trees as output. This method has three key advantages: 1) little hyper-parameter tuning is required, 2) it is computationally efficient, and 3) its ensemble characteristic lessens over fitting. Furthermore, the RF algorithm provides us with some interpretability by way of the ``feature importance'' attribute, with a ranking of features based on their
frequency used as a predictor variable by each tree.

We use the following metrics for scoring the predictive performance of the RF:
\begin{equation}
        \rm{R}^2(y,\hat{y}) = 1 - \frac{\sum_{i=1}^{n}(y_i-\hat{y}_i)^2}{\sum_{i=1}^{n}(y_i-\overline{y})^2},
\end{equation}

\begin{equation}
        \mathrm{RMSE}(y, \hat{y}) = \sqrt{\frac{\sum_{{\rm i}=1}^n(y_{\rm i} - \hat{y}_{\rm i})}{n}},
\end{equation}
where $y_{\rm i}$ are the given target values, $\hat{y}_{\rm i}$ are the RF predicted target values, and $\overline{y}$ is the mean of $y_{\rm i}$. The $\rm{R}^2$ score provides the proportion of the target variable that is predictable by the given features. Because the $\rm{R}^2$ outputs a score between 0.0--1.0, it provides comparable information about performance when comparing various experiments. The RMSE scores, on the other hand, are based on the target value range. Therefore, in order to account for the range in target values across multiple experiments, we normalize our RMSE scores by the Interquartile range (IQR): 
\begin{equation}
     IQR = Q_3 - Q_1, 
\end{equation}
where $Q_3$ is the 3rd quartile (75th percentile) of a given set and $Q_1$ is the first quartile (25th percentile) of the set. In other words, we normalize the RMSE by the middle 50$\%$ dispersion of the target values as RMSE/IQR.

\subsubsection{Features and Targets}
Using the thousands of realizations in the CAMELS {\bf LH} simulation sets, we train a random forest regressor to predict the suppression of the matter power spectrum \DP~as a function of the mean baryon fraction at a range of scales.
We construct the following features for each realization:

\begin{itemize}
    \item \vDBF: the mean baryon fraction of high-mass halos, those with masses $>10^{13.5}$ \Msun
    in each simulation. \\
    \item \BinBF: an array containing mean baryon fraction in 10 bins of halo masses within the range $[10^{10}-10^{15})$ \Msun.\\
    \item \NH: the number of halos within a halo mass range.\\
    \item \NHj: an array with the same shape as \BinBF~containing the number of halos per mass bin.
\end{itemize}

Our target are the \DP~values for each realization at four different $k$-values: $k=[1, 5, 10, 20]\,h$\,Mpc$^{-1}$.

\subsubsection{Robustness of random forest predictions}

One inherent benefit of CAMELS is that we are able to test the effects of feedback model implementation by way of its TNG and SIMBA simulations sets. We create an 80\%\,/\,20\% train/test split of the {\bf LH} simulations and perform the following experiments using either TNG or SIMBA:
\begin{itemize}
\item Train on \vDBF~ to predict \DP~at $k=[1, 5, 10, 20]\,h$\,Mpc$^{-1}$.  We perform this experiment as an extension of vDMS, where in this work we explore a larger halo mass range and probe the non-linear regime.\\
\item Train on \BinBF~at $k=[1, 5, 10, 20]\,h$\,Mpc$^{-1}$.\\
\item Repeat the above two experiments with additional features related to cosmic variance: \NH~or \NHj. 
\end{itemize}

We are further able to test the robustness of our algorithm and determine how well our the RF can marginalize over subgrid physics model by performing ``two-model'' experiments. In these experiments we train on the entire {\bf LH} set of one of the feedback implementation and test on the entire {\bf LH} set of the other, i.e., training on SIMBA and testing on TNG and vice versa. In these experiments we use the same hyperparameters that produced the best results from the previous "single-model" experiments.

%%%%%%%%%%%%%%%%%%%%%%%%%%%%%%%%%%%%%%%%%%%%%%%%%%
\section{Impact of cosmological and feedback parameter variations}
\label{Results: feedback}

In this section we examine the impact of cosmological and baryonic feedback parameter variations on the matter power spectrum and the baryon fraction of halos of different masses, exploring also the connection between the suppression of the total matter power spectrum and the mean baryon fraction of massive halos.
We perform this analysis for both the TNG and SIMBA galaxy formation models. 

\subsection{Matter power spectra}
\label{Results: power spectrum}
We use the total matter power spectra from the {\bf 1P} simulations to examine how cosmological and feedback parameters affect the clustering of matter at various scales. For each run in the {\bf 1P} set, described in section \ref{Methods:CAMELS}, we measure the fractional impact of baryons on the total matter power spectrum, \PhPd. If baryonic physics has no effect on matter clustering, \PhPd~should be of order unity. However, if baryonic physics suppresses the clustering of matter compared to dark-matter only simulations, usually by way of feedback ejecting gas out to large distances, \PhPd~should fall below unity on a range of scales.

\begin{figure*}
    \includegraphics[width=0.99\textwidth]{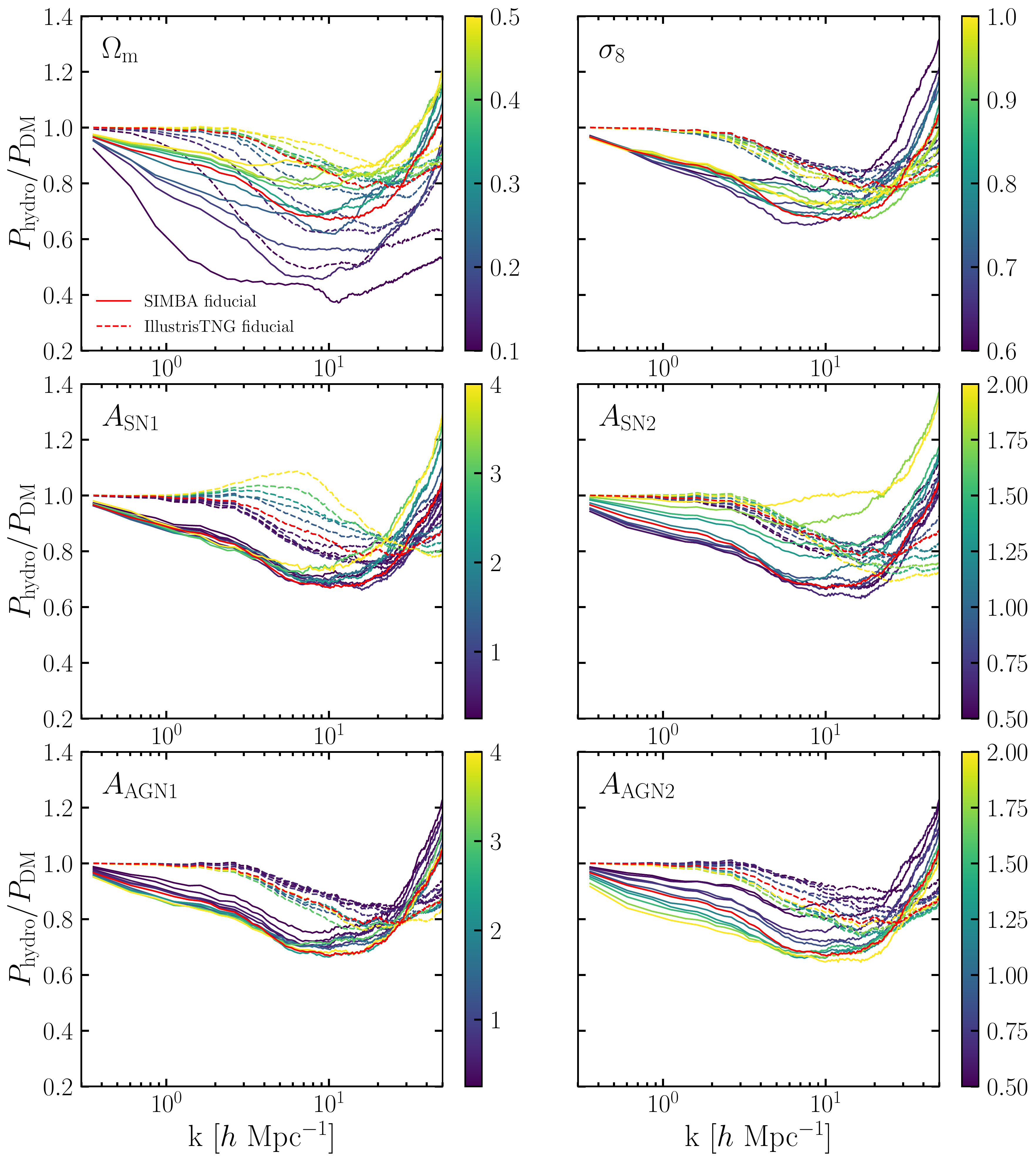}
    \caption{The effect of baryonic physics on matter clustering for different cosmological and feedback parameter variations. 
    Each panel shows the ratio of total matter power spectrum in hydrodynamic simulations to that of the corresponding dark matter-only simulations ($P_{\rm{hydro}}$/$P_{\rm{DM}}$) as a function of wave number $k$ when varying a single parameter in the CAMELS {\bf 1P} sets. Lines of different colors indicate the value of each parameter variation, and red lines indicate the fiducial model for TNG (dashed lines) and SIMBA (solid lines).
    Variations in feedback model, as well as in feedback amplitude, result in variation in total matter clustering. }
    \label{fig:phyd_pdm_1param}
\end{figure*}

Fig.~\ref{fig:phyd_pdm_1param} shows $P_{\rm{hydro}}(k)$/$P_{\rm{DM}}(k)$ as a function of wave number $k$ for the {\bf 1P} simulations, where each spectrum is color coded by the value of each parameter variation.  We notice two overall trends that are roughly independent of cosmological or feedback parameters. The first is the general ``scoop'' shape of $P_{\rm{hydro}}$/$P_{\rm{DM}}$, which is consistent with previous works (\cite{2011MNRAS.415.3649V, 2016MNRAS.461L..11H, 2018MNRAS.474.3173P, 2018MNRAS.480.3962C, 2018MNRAS.475..676S}; vDMS). This shape conveys agreement between matter clustering in hydrodynamical and dark matter-only simulations on large scales ($k \lesssim 0.1\,h$\,Mpc$^{-1}$) while at intermediate scales there is suppression of power by baryonic feedback  ($P_{\rm{hydro}}$/$P_{\rm{DM}} < 1$) and at small scales ($k \gtrsim 40\,h$\,Mpc$^{-1}$) there is enhanced, as opposed to suppressed, clustering relative to dark matter owing to gas dissipative processes ($P_{\rm{hydro}}$/$P_{\rm{DM}} > 1$). 
The second overall trend is that the SIMBA galaxy formation model (solid lines) tends to suppress power on intermediate scales more strongly compared to the TNG galaxy formation model (dashed lines) while driving a steeper increase in small-scale clustering ($k \lessapprox30\,h$\,Mpc$^{-1}$). 

We now analyze in more detail how each parameter affects matter clustering by comparing $P_{\rm{hydro}}(k)$/$P_{\rm{DM}}(k)$ between the fiducial models of TNG and SIMBA (shown in red) and that of the individual parameter variations:
\begin{itemize}

\item \textit{Cosmological parameters}: The top two panels in Fig.~\ref{fig:phyd_pdm_1param} show the sensitivity of the total matter power spectrum to $\Omega_{\rm m}$ and $\sigma_8$ for a fixed galaxy formation model. We see a strong dependence of $P_{\rm{hydro}}$/$P_{\rm{DM}}$ on the value of $\Omega_{\rm m}$ both in TNG and SIMBA. As $\Omega_{\rm m}$ decreases (at fixed $\Omega_{\rm b}$), there is a greater suppression of power on intermediate scales. This can be understood as a consequence of baryons contributing a higher fraction of the total matter content making feedback more efficient at pushing gas out of halos and distributing matter on larger scales, in agreement with the analysis of large-scale baryon spread in \citet{2023arXiv230711832G}. 
In contrast, we identify weaker trends for $\sigma_8$, with significant scatter. \\

\item \textit{Supernova feedback parameters}: The middle two panels in Fig.~\ref{fig:phyd_pdm_1param} show the impact of changing the stellar feedback parameters $A_{\rm{SN1}}$ and $A_{\rm{SN2}}$ respectively, which control the mass loading and velocity of galactic winds, on matter clustering. Both panels show somewhat counterintuitive effects of stellar feedback.  Increasing $A_{\rm{SN1}}$ in SIMBA reduces (rather than enhances) the suppression of power on small scales ($k \gtrsim 10\,h$\,Mpc$^{-1}$) and increasing $A_{\rm{SN2}}$ further increases $P_{\rm{hydro}}$/$P_{\rm{DM}}$ over the full range of scales. This can be understood as a consequence of the nonlinear interplay between stellar and AGN feedback, where stronger stellar feedback suppresses black hole growth and results in weaker effective impact of AGN feedback on matter clustering \citep{2011MNRAS.415.3649V,2023arXiv230102186P}.  The TNG model shows rather different trends, with reduced suppression of power on scales $k \lesssim 10$--20$\,h$\,Mpc$^{-1}$ but enhanced suppression of power on smaller scales when increasing $A_{\rm{SN1}}$ and $A_{\rm{SN2}}$.
These results are consistent with the analysis of electron power spectra in CAMELS by \cite{2022JCAP...04..046N}, which highlights the sensitivity of predicted baryonic effects on galaxy formation implementation.  \\

\item \textit{AGN feedback parameters}: The bottom two panels of Fig.~\ref{fig:phyd_pdm_1param} show the impact of varying AGN feedback efficiency on matter clustering. In this case, there are clear systematic trends for stronger suppression of power when increasing both $A_{\rm{AGN1}}$ and $A_{\rm{AGN2}}$ for both galaxy formation models (TNG and SIMBA).
The sensitivity of $P_{\rm{hydro}}$/$P_{\rm{DM}}$ to $A_{\rm{AGN1}}$ is weaker given its range of variation, with no more than 10\% difference relative to the fiducial model. In contrast, the matter power spectrum in SIMBA displays a strong sensitivity to the AGN jet speed, $A_{\rm{AGN2}}$, with strong suppression of power across scales, reaching $P_{\rm{hydro}}$/$P_{\rm{DM}} \sim 0.6$ at $k \sim 10\,h$\,Mpc$^{-1}$ with jets twice as fast relative to the fiducial model.
This results are also consistent with previous findings for electron power spectra \citet{2022JCAP...04..046N} and the impact of large scale jets on cosmological baryon spread \citep{2023arXiv230711832G}. 
\end{itemize}

\subsection{Halo baryon fraction}
\begin{figure*}
    \includegraphics[width=0.99\textwidth]{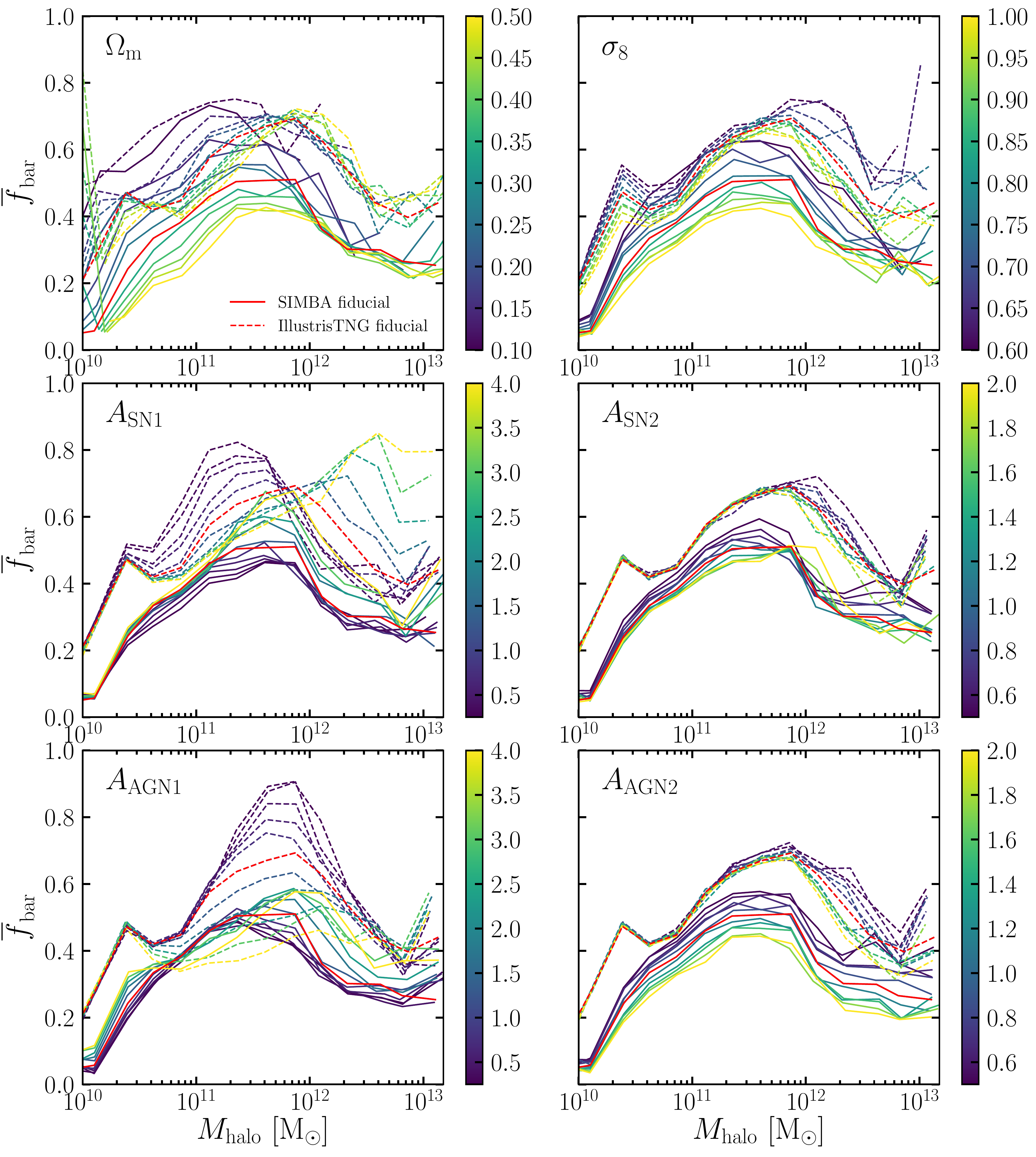}
    \caption{The effect of each parameter variation on the mean halo baryon fraction as a function of halo mass, \BF(M$_{\rm{halo}}$). 
    As in Fig.~\ref{fig:phyd_pdm_1param}, the color bar for each panel indicates the value of the corresponding parameter variation while all other parameters are held constant. 
    The red lines indicate the fiducial parameters for TNG (dashed lines) and SIMBA (solid lines), with their peak baryon fraction occurring at $\lesssim 10^{12}$ \Msun~in both fiducial models. The \BF~values and their halo mass dependence differ substantially between galaxy formation implementations and model parameter variations.
    }
    \label{fig:1P_bf_vs_vmassom}
\end{figure*}

In Fig.~\ref{fig:1P_bf_vs_vmassom} we use again the specialized CAMELS {\bf 1P} simulation sets to analyze the impact of individual cosmological and feedback parameter variations on the average halo baryon fraction as a function of halo mass, \BF($M_{\rm{halo}}$), where we consider logarithmically-spaced halo mass bins in the range $10^{10}$--$10^{13}$ \Msun. 
We notice two main trends roughly independent of cosmological or feedback parameters when comparing the fiducial realizations (indicated in red) for the TNG (dashed lines) and SIMBA (solid lines) models.
The first being that peak of the halo baryon fraction occurs at $\sim 10^{12}$ \Msun. We notice a drop in mean baryon fraction as halos exceed this mass range, when powerful feedback process can expel material out of the halo. We note, however, that at very high mass halos, we expect feedback to be less efficient at expelling material and for there to be another rise in mean baryon fraction.
The second main trend is that SIMBA has overall lower \BF($M_{\rm{halo}}$) compared to TNG, with the fiducial models reaching their peak at \BF$\sim0.5$ for SIMBA and \BF$\sim0.7$ for TNG.

We now analyze in more detail how each CAMELS parameter variation affects halo baryon fractions, keeping in mind that the definitions of feedback parameters are not the same for TNG and SIMBA:
\begin{itemize}

\item \textit{Cosmological parameters}: The top two panels of Fig.~\ref{fig:1P_bf_vs_vmassom} show the sensitivity of \BF($M_{\rm{halo}}$) to our cosmological parameters $\Omega_{\rm m}$ and $\sigma_8$.  
Halo baryon fractions appear to be more sensitive to cosmology in SIMBA compared to TNG.  However, both galaxy formation models predict qualitatively similar trends, with lower \BF($M_{\rm{halo}}$) when increasing $\Omega_{\rm m}$ and $\sigma_8$ across a range of halo masses.
This trend may seem trivial for $\Omega_{\rm m}$ since we hold the value of $\Omega_{\rm{b}}$ constant in all CAMELS simulations, implying that the average cosmic baryon fraction decreases with higher $\Omega_{\rm m}$ and so should the corresponding halo baryon fractions.  However, \BF($M_{\rm{halo}}$) is normalized by $\Omega_{\rm b}/ \Omega_{\rm m}$ for each simulation (Equation~\ref{eq:fbar}), removing the trivial effect of varying $\Omega_{\rm m}$ at fixed $\Omega_{\rm b}$. The impact of increasing $\Omega_{\rm m}$ on \BF($M_{\rm{halo}}$) is thus a reflection of the effective efficiency of feedback when changing the amount of baryons relative to the dark matter gravitational potential, and this effect seems more prominent in lower mass halos for both TNG and SIMBA. Interestingly, the baryon fraction decreases systematically at all halo masses when increasing $\sigma_8$, while the suppression of power does not seem to follow a clear trend with $\sigma_8$.  \\

\item \textit{Supernova feedback parameters}: The middle two panels in Fig.~\ref{fig:1P_bf_vs_vmassom} show the impact of changing $A_{\rm{SN1}}$ and $A_{\rm{SN2}}$ on halo baryon fractions. As for the power spectra, varying the mass loading of galactic winds ($A_{\rm{SN1}}$) has a different effect in each galaxy formation model. We might intuitively expect that as $A_{\rm{SN1}}$ increases, more gas would be ejected out of galaxies resulting in lower \BF($M_{\rm{halo}}$). However, we only see this behavior in TNG for halos with mass $M_{\rm{halo}} \lesssim 10^{12}$ \Msun, while the baryon fraction of higher mass halos increases with $A_{\rm{SN1}}$ owing to the suppression of AGN feedback. This reversed trend with $A_{\rm{SN1}}$ can explain its impact on $P_{\rm{hydro}}$/$P_{\rm{DM}}$ for TNG, with an enhancement of power on intermediate scales coinciding with the highest gas fraction in high mass halos with $A_{\rm{SN1}} = 4$.
Meanwhile, SIMBA shows systematically higher baryon fractions when increasing $A_{\rm{SN1}}$ across the full halo mass range, indicating a different non-linear coupling of stellar and AGN feedback compared to TNG, which can explain the suppression of power seen in Fig.~\ref{fig:phyd_pdm_1param}.
On the other hand, increasing the speed of galactic winds ($A_{\rm{SN2}}$) results in systematically lower \BF($M_{\rm{halo}}$) values for both SIMBA and TNG, but in this case lower baryon fractions correlate with less suppression of matter clustering in SIMBA on all scales. \\

\item \textit{AGN feedback parameters}: The bottom two panels of Fig. \ref{fig:1P_bf_vs_vmassom} show the sensitivity of \BF($M_{\rm{halo}}$) to changes in AGN feedback efficiency.  
Halo baryon fractions are significantly reduced by increasing the kinetic mode black hole feedback efficiency $A_{\rm{AGN1}}$ in TNG in the intermediate halo mass range $M_{\rm{halo}} = 10^{11}$--$10^{12.5}$ \Msun, while the burstiness parameter $A_{\rm{AGN2}}$ has a stronger effect reducing \BF($M_{\rm{halo}}$) in higher mass halos. In both cases, the decrease in halo baryon fraction with higher AGN feedback efficiency correlates with stronger suppression of matter clustering.
Similarly, increasing the AGN jet speed in SIMBA ($A_{\rm{AGN2}}$) drives an overall reduction of halo baryon fractions and increased suppression of matter clustering on all scales shown in Fig.~\ref{fig:phyd_pdm_1param}, corresponding to more efficient spread of baryons on large scales relative to the TNG model \citep{Tillman2022_Simba_Lya,2023arXiv230711832G}. However, the effect of increasing the momentum flux $A_{\rm{AGN1}}$ in SIMBA seems more complex, driving an increase in baryon fraction in high mass halos (possibly due to black hole self-regulation) but stronger suppression in the matter power spectrum, particularly at low k-values.  
\end{itemize}

\subsection{Suppression of matter power spectrum as a function of baryon fraction}

Using a suite of matter power spectra from hydrodynamical and dark matter only simulations, vDMS found a tight relation between the suppression of the matter power spectrum (\DP; defined in \S\ref{Methods: variables}) in the linear regime and the average baryon fraction (\BF) of high mass halos ($\sim 10^{14}$ \Msun). 
In this study, a direct comparison to vDMS is not possible due to the small size of the individual CAMELS realizations and relatively small number of high mass halos. We therefore extend what was done in vDMS and use the {\bf LH} simulation sets of CAMELS, described in \S\ref{Methods:CAMELS}, to investigate how \DP~is affected by cosmological and baryonic feedback parameters over a broader range of model variations.

\begin{figure*}
    \centering
    \includegraphics[width=0.99\textwidth]{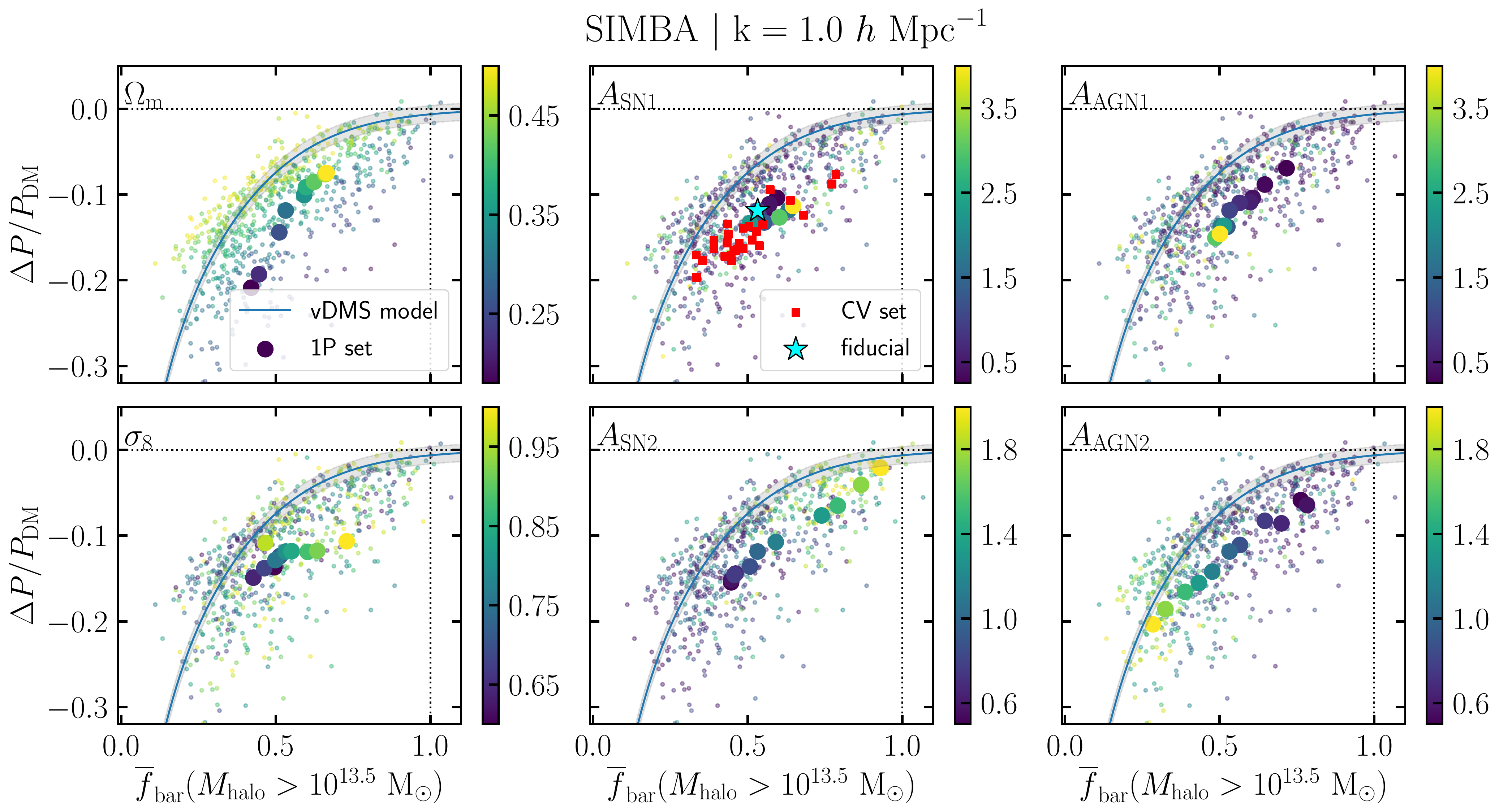}
    \caption{Suppression of the matter power spectrum, \DP, as a function of mean baryon fraction of high mass halos, \vDBF.
    The blue line is the fitting function of vDMS for a halo definition of 200c (200 times the critical density of the Universe), with the gray shaded region indicating 1\%~variation in \DP. Each panel shows the {\bf 1P} set overlaid on the {\bf LH} set, color coded by the value of each of the six parameters. We remind the reader that all six parameters are varied simultaneously in the {\bf LH} set, while only one parameter is varied in the {\bf 1P} set.
    The middle top panel shows additional results from the {\bf CV} simulation set, where all six parameters are constant and only the initial conditions are varied. The fiducial realization is indicated by the cyan star.
    We find that \DP~increases for higher values of \BF, meaning that there is less suppression of the matter power spectrum in simulations where feedback is less effective at removing gas from halos. Data points fall generally along the vDMS model, with large scatter owing to broad parameter variations and cosmic variance.
    }
    \label{fig:SIMBA_VDplots}
\end{figure*}

\begin{figure*}
    \includegraphics[width=0.99\textwidth]{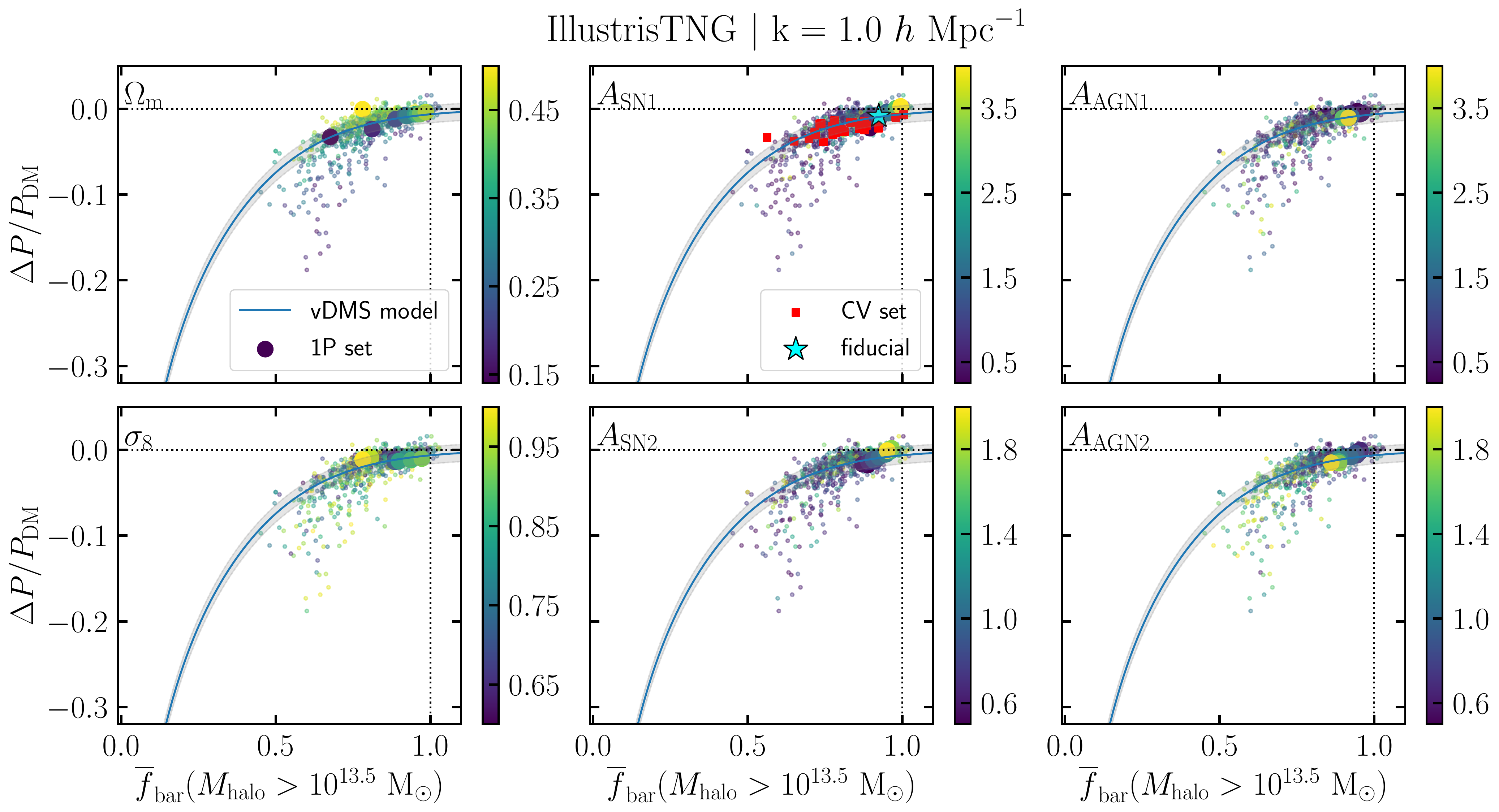}
    
    \caption{Same as Fig. \ref{fig:SIMBA_VDplots} but for the TNG {\bf LH} and {\bf 1P} simulation sets. The TNG sets produce a smaller range of variation in \DP~and \BF~compared to the SIMBA sets and are in closer agreement with the vDMS model than SIMBA. The location of our TNG fiducial realization (cyan star) along the vDMS model is consistent with the location of the original IllustrisTNG simulations in the vDMS study.}
    \label{fig:TNG_VDplots}
\end{figure*}

Fig.~\ref{fig:TNG_VDplots} and Fig.~\ref{fig:SIMBA_VDplots} show 
$\Delta P/{P_\mathrm{DM}} \equiv (P_\mathrm{hydro}-P_\mathrm{DM})/P_\mathrm{DM}$ evaluated at $k = 1.0\,h$\,Mpc$^{-1}$ as a function of \BF~for the TNG and SIMBA {\bf 1P} and {\bf LH} sets, respectively.
We note that due to the small simulated volumes in CAMELS there are not enough halos of mass $\sim 10^{14}$ \Msun~to replicate the results in vDMS
and we therefore evaluate \BF~for halos with mass $>10^{13.5}$ \Msun.
Each panel reproduces the same data points depicting the {\bf LH} sets (small dots), overlaid by the {\bf 1P} set (large circles) corresponding to the labeled parameter in that panel. The data are color coded by the parameter value. We remind the reader that each of the six parameters are simultaneously varied in the {\bf LH} sets, while only one parameter is varied in the {\bf 1P} sets. We thus examine how individual cosmological and feedback parameters affect the relation between \DP~and \BF, which we compare to the fitting function derived by vDMS (their equation 5) for baryon fractions calculated using the 200c virial definition, which we henceforth refer to as the ``vDMS model'' and indicate by the blue solid line and gray shaded region.
Lastly, we overlay the results for the {\bf CV} sets of TNG and SIMBA (described in \S\ref{Methods:CAMELS}) as the red squares in the top center panel of each figure in order to examine the effect of cosmic variance on this relation. We emphasize that this represents a lower limit of cosmic variance as the 27 {\bf CV} realizations probe only a small volume. We have indicated where the fiducial realization (cyan star) lies within the spread. It is interesting to note that there is significant spread due to cosmic variance in both \DP~and \BF~in SIMBA compared to the fiducial run.

Our CAMELS results in Figs.~\ref{fig:SIMBA_VDplots} and~\ref{fig:TNG_VDplots} reveal a good qualitative agreement with the general trend found in vDMS: the suppression of the matter power spectrum increases as the average baryon fraction in massive halos decreases. We note that we have kept the y-axis limits the same in both figures for a more clear comparison of TNG and SIMBA, however, there are several SIMBA data points in Fig. \ref{fig:SIMBA_VDplots} that fall below the visible y-axis. While the SIMBA {\bf LH} set probes a range of \DP~and \BF~values significantly larger than the TNG {\bf LH} set, as expected from Figs.~\ref{fig:phyd_pdm_1param} and~\ref{fig:1P_bf_vs_vmassom}, both models roughly follow the vDMS trend, suggesting that \BF~in massive halos can be used to infer the redistribution of baryons over large scales regardless of galaxy formation model implementation. The location of 
 the TNG fiducial run is in strong agreement with the vDMS model. The fiducial run for SIMBA, however, appears to fall outside of the vDMS fit line, and is further discussed in section \ref{summary} of this paper.
However, we find considerable spread in \DP~at fixed \BF~compared to vDMS, which can be attributed to the broader range of parameter variations explored in CAMELS.:
\begin{itemize}

\item \textit{Cosmological parameters}: The left two panels in Figs.~\ref{fig:SIMBA_VDplots} and~\ref{fig:TNG_VDplots} explore the dependence of the \DP--\BF~relation on $\Omega_{\rm m}$ and $\sigma_8$. We remind the reader that \BF~is normalized by $\Omega_{\rm b}$/ \Om.
For both TNG and SIMBA, the {\bf LH} sets show a trend of higher \DP~(i.e. less suppression of power) at fixed \BF~for higher values of $\Omega_{\rm m}$ in high-mass halos. This implies that the same impact on the total matter power spectrum (at $k = 1.0\,h$\,Mpc$^{-1}$) can be predicted by simulations that yield different halo baryon fractions, in this case as a consequence of the different response of feedback to changes in $\Omega_{\rm m}$ at fixed $\Omega_{\rm b}$. We further note that the {\bf LH} data points which stray furthest from the vDMS model correspond to the lower end values of the \Om~variation. The SIMBA {\bf 1P} set more clearly shows the trend of decreased suppression of the matter power spectrum as a function of mean baryon fraction with increased value of \Om~in high-mass halos.

There is also a visible, albeit less pronounced, trend for $\sigma_8$, where 
\DP~becomes more negative (i.e. stronger suppression of power) at fixed \BF~for higher values of $\sigma_8$.
Overall, the non-linear response of the fiducial galaxy formation model to variations in cosmology appears to explain a significant fraction of the scatter in the vDMS relation seen for SIMBA and TNG. \\ 

\item \textit{Supernova feedback parameters}: The middle panels of Figs.~\ref{fig:SIMBA_VDplots} and~\ref{fig:TNG_VDplots} explore the dependence of the vDMS relation on systematic variations of the mass loading factor and speed of galactic winds driven by stellar feedback (parameters $A_{\rm{SN1}}$ and $A_{\rm{SN2}}$, respectively).
In TNG, there is indication for simulations clustering around \DP$\,\sim 0$ and \BF$\sim 1$ for higher values of $A_{\rm{SN1}}$ and $A_{\rm{SN2}}$, corresponding to weaker overall impact of feedback owing to the suppression of black hole growth and therefore AGN feedback. We find qualitatively similar trends in SIMBA for variations in $A_{\rm{SN2}}$, with a clear trend in the {\bf 1P} set showing weaker suppression of the power spectrum and higher \BF~with the increase of galactic wind speed due to stellar feedback.
Decreasing the strength of stellar feedback parameters (indicated by dark purple points) tends to yield more negative \DP~values and correspondingly lower \BF~(i.e., stronger impact). This displaces realizations roughly along the vDMS relation but with increasing scatter.\\

\item \textit{AGN feedback parameters}: The right two panels of Figs.~\ref{fig:SIMBA_VDplots} and~\ref{fig:TNG_VDplots} show the impact of AGN feedback parameter variations in the vDMS relation. In this case, the large range of both \DP~and \BF~values in SIMBA allows for a clear depiction of the  stronger dependence in AGN parameters along the vDMS relation. The SIMBA {\bf 1P} set in particular shows that stronger AGN feedback leads to both lower values of \DP~and lower \BF, indicating more efficient evacuation of gas from halos and stronger suppression of total matter clustering. The trend is also present in TNG, more so in $A_{\rm{AGN2}}$ than in $A_{\rm{AGN1}}$ (as expected from Figs.~\ref{fig:phyd_pdm_1param} and~\ref{fig:1P_bf_vs_vmassom}), albeit less clear due to the tight assembling of the TNG data points along a smaller range of the vDMS model.\\

\item \textit{Cosmic variance}: In order to examine how cosmic variance affects the predicted variation of \DP~as a function of \BF, the top middle panel of Figs.~\ref{fig:SIMBA_VDplots} and~\ref{fig:TNG_VDplots} overlay the results from the CAMELS {\bf CV} sets corresponding to 27 realizations of the fiducial TNG and SIMBA models using different initial conditions (red triangles). For TNG, the {\bf CV} set yields roughly similar range in average baryon fraction of massive halos as the entire {\bf LH} set, indicating that stochastic variations owing to the small CAMELS volumes and correspondingly low number of massive halos play an important role.  Nonetheless, the TNG CV simulations roughly follow the vDMS relation.  The SIMBA CV set also yields a wide range of \DP~and \BF~values, but in this case suggesting a systematic offset relative to the vDMS model.

\end{itemize}

We can quantify the impact of cosmic variance on the predicted suppression of matter clustering as the root mean square variation in \DP~relative to the mean: 
\begin{equation}\label{eq:dispersion}
    \delta_{\rm cv} \equiv \frac{\sigma_{\rm cv}}{\left|\overline{p_{\rm{cv}}}\right|},
\end{equation}
with $p_{\rm{cv}} \equiv$~\DP~ for the {\bf CV} set and
\begin{equation}
    \sigma_{\rm cv}^2 = \rm{\frac{1}{n} \sum_{i=1}^{n} \left( p^{i}_{\rm{cv}}- \overline{p_{\rm{cv}}}\right)^2},
\end{equation}
where $n=27$ realizations and $\overline{p_{\rm{cv}}}$ represents the average of $p_{\rm{cv}}$ over the {\bf CV} set.
Evaluating Eq.~\ref{eq:dispersion} for $k = 1.0\,h$\,Mpc$^{-1}$ gives $\delta_{\rm cv} = 0.192$ for SIMBA, and $\delta_{\rm cv} = 0.357$ for TNG, indicating that there is considerable variation due to cosmic variance alone.

The considerable spread of CAMELS predictions relative to the vDMS model shown in this section provides motivation for the machine learning experiments described in section \ref{Methods :ML}. Given the larger data set in CAMELS with broader variations in feedback and cosmology compared to previous libraries of power spectra, it is possible that the vDMS model relating halo baryon fraction and suppression of matter clustering is not general enough to include every plausible feedback model. For example, we later examine the original SIMBA model against the vDMS relation and find that SIMBA does not fall within 1\% of the vDMS fit, as shown in Fig.\ref{fig:vDSIMBA}. 
However, it is also possible that having smaller volumes which are significantly affected by cosmic variance as compared to the data set in vDMS, along with the lack of halos of mass $\sim 10^{14}$ \Msun, may explain the disagreement between our results and the vDMS model. These results motivate us to explore the relation between \DP~and halo baryon fraction with a machine learning approach, where we can extract information from a broader halo mass range to improve the accuracy of predictions for the impact of baryonic physics on the total matter power spectrum.

\section{Estimating the impact of feedback on matter clustering with machine learning}
\label{Results:ML}

A major goal of this work is to show that machine learning can be used to extract information from the full range of halo masses in order to estimate the suppression of the matter power spectrum by baryonic processes all the way to the non-linear regime. In this section, we discuss the results of training a random forest regressor (RF) to estimate the impact of feedback on the clustering of matter using the {\bf LH} simulation sets in CAMELS, which vary simultaneously cosmological and feedback parameters (\S\ref{Methods:CAMELS}).
The general setup of our experiments is described in \S\ref{Methods :ML}.

\begin{figure*}
    \centering
    \includegraphics[width=0.99\textwidth]{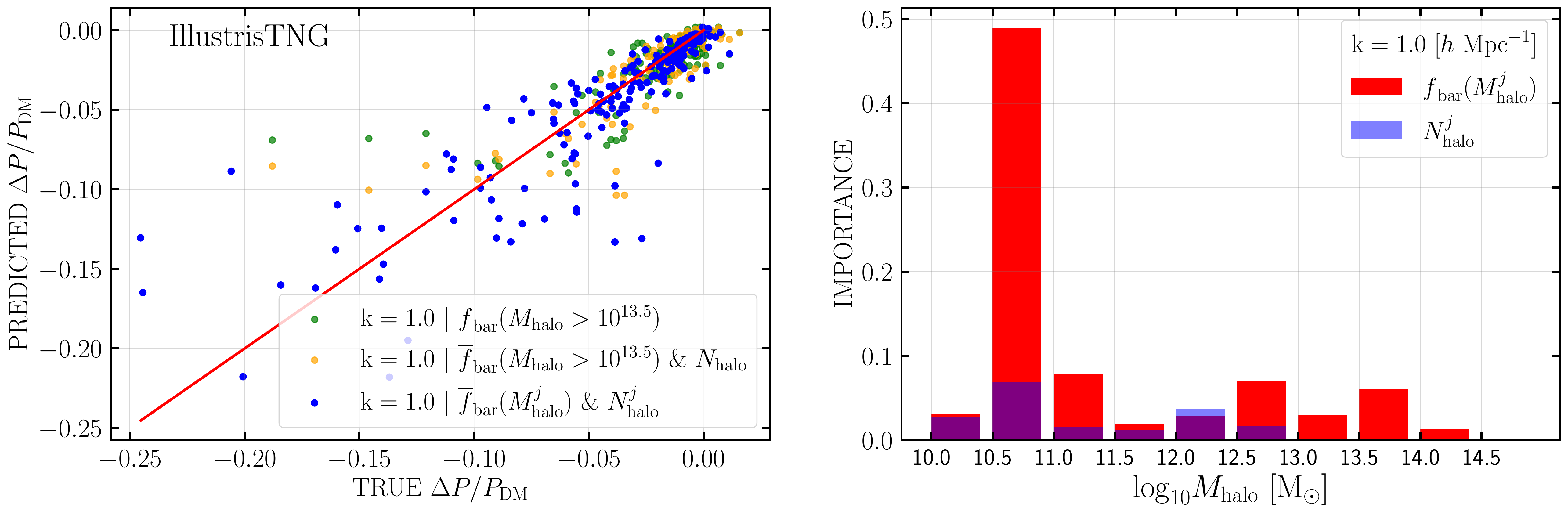}
    \includegraphics[width=0.99\textwidth]{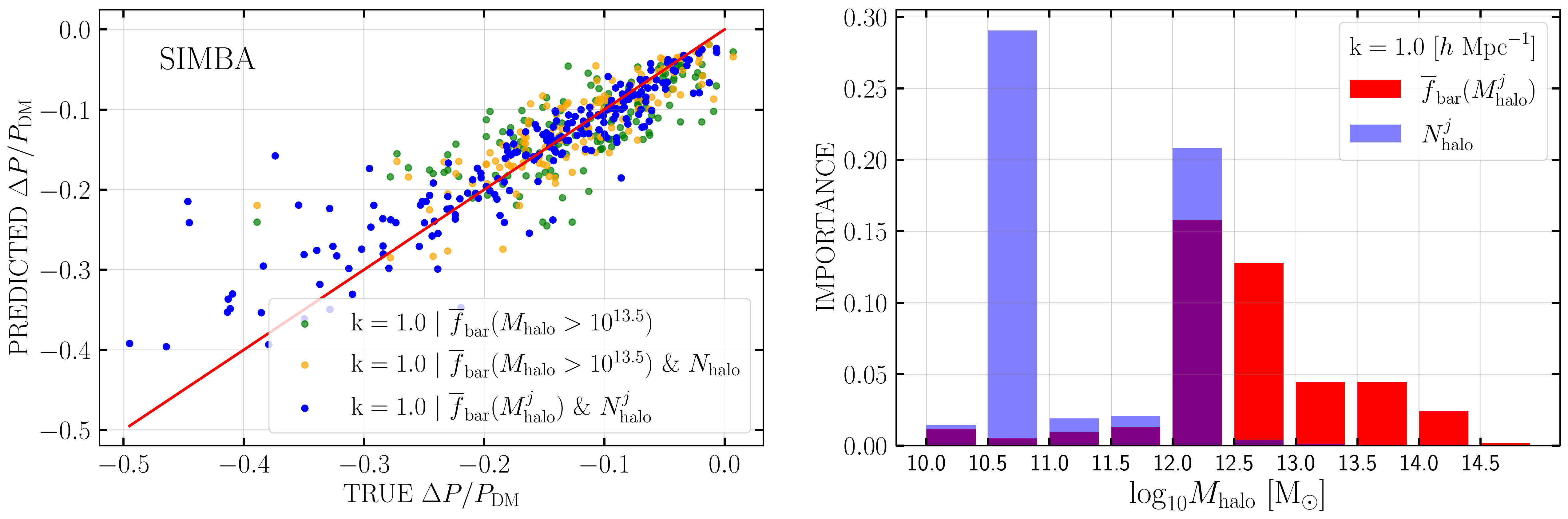}
    \caption{Results from different Random Forest experiments. We create an 80/20 train/test split of the {\bf LH} data sets to predict \DP~at $k=1.0$ for TNG (top) and SIMBA (bottom). The \textit{left} panels show the predicted target values compared to the true target values as given by the test sets in CAMELS, where the red line indicates a perfect one-to-one relation. The green data points correspond to predictions by a RF trained only on the baryon fraction of massive halos, \vDBF, the orange data points show results for a RF trained on \vDBF~as well as the number $N_{\rm{halo}}$ of massive halos, while results from training a RF on \BinBF~and \NH~ for halo mass bins $j$ spanning the full mass range are shown in blue. The \textit{right} panels show the feature importances corresponding to the blue data points in the left panel (training on \BinBF~and \NH), indicating the relative rank ordering of importance (from 0.0 to 1.0) given to each training feature by the RF. 
    Predictions improve by providing training data across the full range of halo masses.
    }
    \label{fig:RF_vDMS_vs_binned}
\end{figure*}

\begin{figure*}
    \centering
    \includegraphics[scale=0.4]{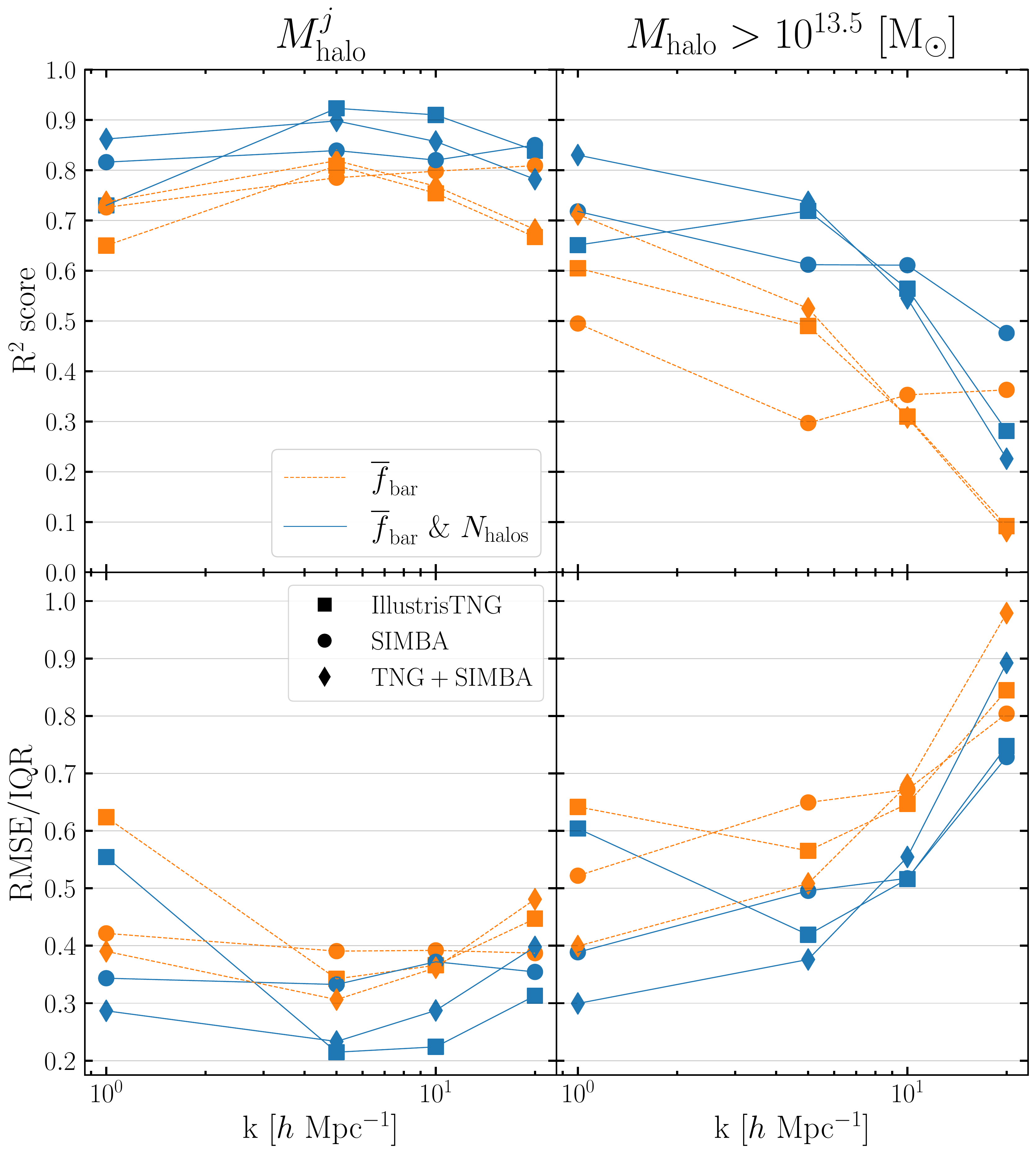}
    \caption{Performance scores for all RF experiments predicting \DP~at four different scales, $k=[1, 5, 10, 20]\,h$\,Mpc$^{-1}$. The orange dashed lines connect data points showing scores for a RF trained on average baryon fractions \BF, while the blue solid lines connect those where both \BF~and the number of halos $N_{\rm{halo}}$ are used as training features. We show results for three different data sets: the TNG {\bf LH} set (squares), the SIMBA {\bf LH} set (circles), and the two {\bf LH} sets combined, labeled ``TNG+SIMBA'' (diamonds). Left panels correspond to training on features from a range of halo masses, \BinBF, while the right panels correspond to training on high-mass halos only, $M_\mathrm{halo} > 10^{13.5}$ \Msun, as described in section \ref{Methods :ML}.
    We report $\rm{R}^2$ scores ({\it top}) and RMSE scores normalized by the interquartile range (IQR) of the respective data set ({\it bottom}); higher $\rm{R}^2$ scores and lower RMSE/IQR scores convey an improved performance.    
    We obtain higher scores when training on \BinBF~and \NH~for a range of halo masses at all scales compared to training on high-mass halos only, and the highest scores occur in the highly non-linear regime at $k\sim 5$--10$\,h\,{\rm Mpc}^{-1}$, meaning that important information can be extracted from a range of halo masses in the non-linear regime. 
     }
    \label{fig:scores}
\end{figure*}

\subsection{Extracting information across the halo mass range with random forest regression}

Fig.~\ref{fig:RF_vDMS_vs_binned} shows the results from training a RF regressor on different input features to estimate the suppression of power \DP~at $k = 1.0\,h$\,Mpc$^{-1}$.  
We begin by training a RF with \vDBF~as the only training feature, in analogy with the information used by the vDMS fitting function. Halos of mass $\sim 10^{13.5}$ \Msun~are only available for $\sim$700 out of 1,000 {\bf LH} realizations for each of TNG and SIMBA, limiting the size of the training set.
In this first experiment, the RF is only able to predict $\sim$60\%~and $\sim$50\%~of the variation of \DP~in TNG and SIMBA, respectively, with the predicted versus true values of \DP~shown by the green data points in the left panels of Fig. \ref{fig:RF_vDMS_vs_binned}. 
Next, we add the number of high mass halos $N_{\rm{halo}}$ corresponding to the measured \vDBF~as an additional input feature, with results indicated by the orange data points. In this case, the RF predictions improved by $\sim$5\%~in TNG and $\sim$20\%~for SIMBA. 
We then incorporate information from halos across the full mass range by introducing the baryon fraction \BinBF~and the corresponding number of halos \NH~within each halo mass bin (see \S\ref{Methods :ML}), with results shown by the blue data points. In this case, we can use the full {\bf LH} sets of CAMELS for training and testing since we are not limited by the availability of high mass halos. With these additional features using information from a range of halo masses, the RF predicted $\sim$75$\%$ of the variation in \DP~at $k = 1.0\,h$\,Mpc$^{-1}$ in TNG and $\sim$82$\%$ SIMBA,  significantly improving upon the original results.

As stated in section \ref{Methods :ML}, one advantage of the RF is that it provides some level of interpretability by means of the ``feature importance'' attribute. The right panels of Fig.~\ref{fig:RF_vDMS_vs_binned} display the relative importance assigned to each feature by the trained RF on the test set for TNG (top) and SIMBA (bottom). We stress that this ranking is based on the frequency with which the features are used by each tree in order to predict the target variable, and it is possible for more than one feature to hold similar, correlated information but be ranked differently by the RF. Interestingly, the RF ranked \BF~in halos with mass $M_{\rm halo} = 10^{10.5}$--$10^{11.0}$ \Msun~as the most important feature to predict the suppression of power \DP~at $k = 1.0\,h$\,Mpc$^{-1}$ in TNG. Furthermore, we see that the RF ranked several features across the mass ranges as important predictors of \DP~in SIMBA. These feature importance results reveal that the RF was able to extract valuable information across a range of halo masses.

\subsection{Random forest predictions in the highly non-linear regime}

In the previous subsection, we have established that training the RF on features from a range of halo masses improves the predictions for \DP~in the linear regime compared to using \vDBF~as a training feature alone. We now exploit the same methodology to extend our predictions into the highly non-linear regime by repeating our experiments at a range of scales, predicting \DP~at $k=[1, 5, 10, 20]\,h$\,Mpc$^{-1}$.
Fig.~\ref{fig:scores} provides a summary of performance scores for these experiments using the {\bf LH} simulation sets for TNG (squares) and SIMBA (circles). We also performed additional experiments using the two {\bf LH} data sets combined, which we refer to as ``TNG+SIMBA'' (diamonds). 
Descriptions of the scoring metrics can be found in \S\ref{Methods :ML}. 
The top panels in Fig. \ref{fig:scores} show the $\rm{R}^2$ scores and the bottom two panels show the RMSE scores normalized by the interquartile range (IQR). We normalize the RMSE by the IQR in order to account for the variation in the range of \DP, which depends on the $k$ value; the range of \DP~increases as we move to non-linear regimes. 
In addition to presenting results at a range of scales for each training set, Fig.~\ref{fig:scores} compares the predictions based on \vDBF~alone (right panels) versus providing the baryon fraction \BinBF~in different halo mass bins (left panels).  Results based on baryon fractions alone are shown in orange while results that also incorporate the corresponding number of halos \NH~are shown in blue.

\begin{figure*}
    \centering
    \includegraphics[width=0.99\textwidth]{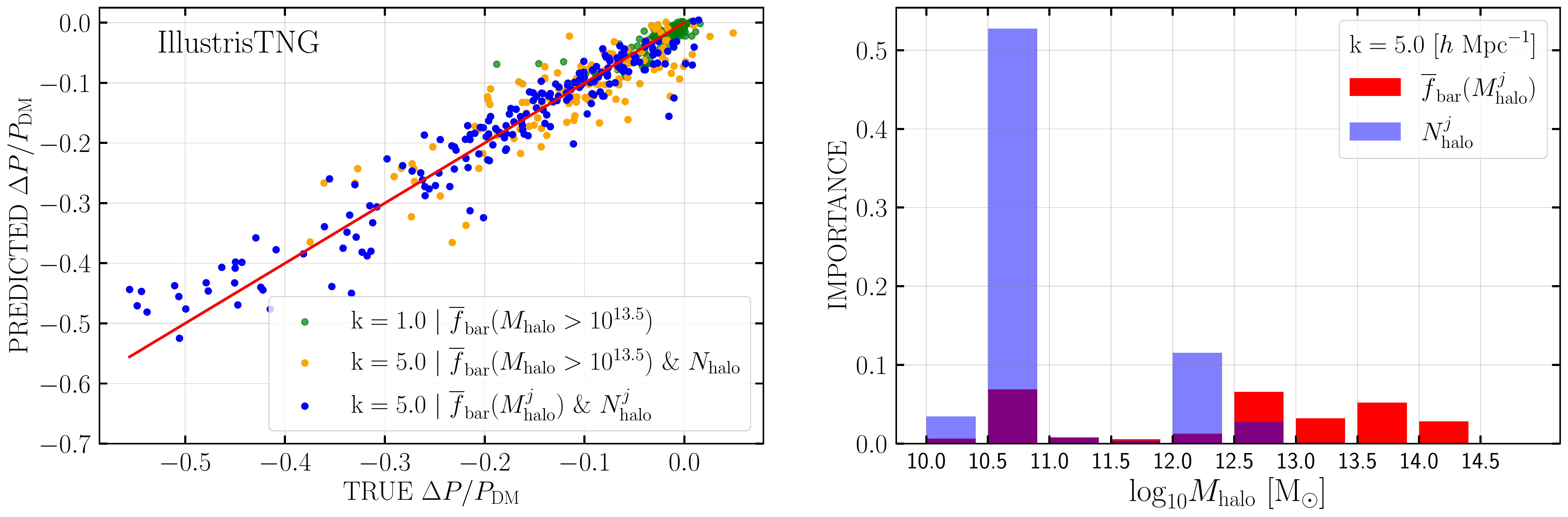}
    \includegraphics[width=0.99\textwidth]{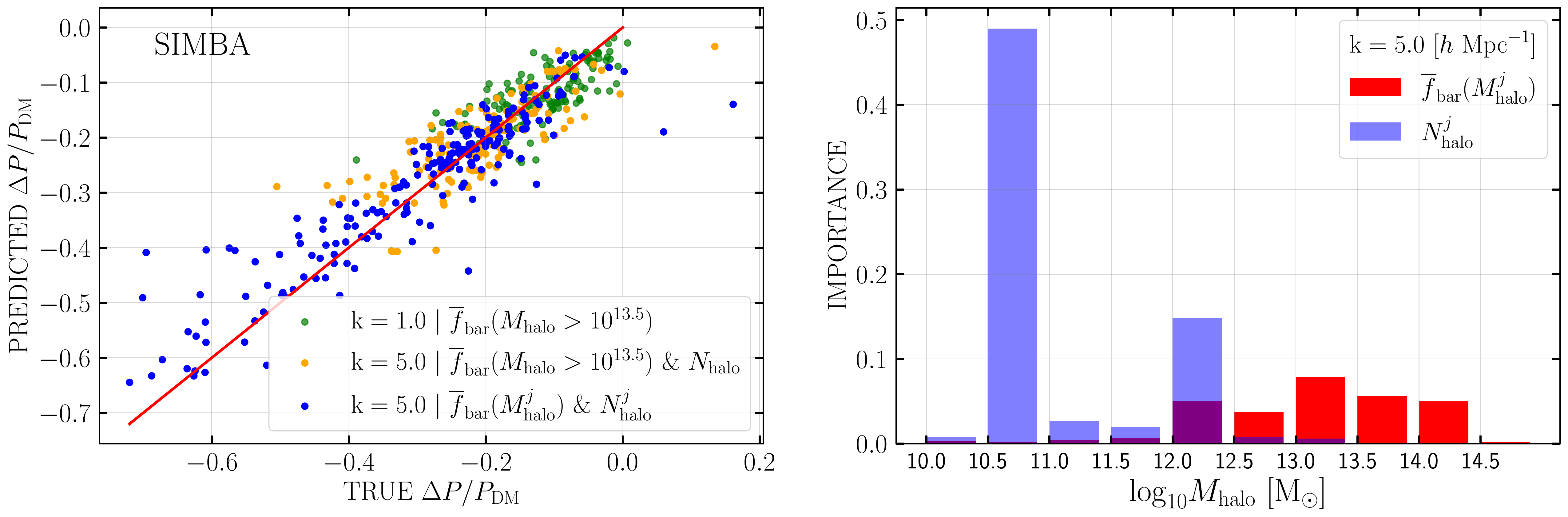}   
    \caption{Similar to Fig. \ref{fig:RF_vDMS_vs_binned} but for results at $k=5.0\,h$\,Mpc$^{-1}$. The \textit{left} panels show the predicted versus true \DP~values for RF regressors trained on \vDBF~and $N_{\rm{halo}}$ (orange) and trained on \BinBF~and \NHj~(blue), both at $k = 5\,h$\,Mpc$^{-1}$. For comparison, we also show the results using only \vDBF~at $k=0.5\,h\,\mathrm{Mpc}^{-1}$ as in Fig.~\ref{fig:RF_vDMS_vs_binned} (green). The \textit{right} panels show the feature importances when training on \BinBF~and \NHj.
    Going to non-linear regimes the range of the target value \DP~increases, enabling more accurate predictions.   
    The highest ranked feature at $k=5.0\,h$\,Mpc$^{-1}$ is \NHj~at $M_{\rm halo} = 10^{10.5}$--$10^{11}$ \Msun, meaning that the number of low-mass halos is highly informative for estimating the power at 1Mpc length scales.}
    \label{fig:RF_k5_plots}
\end{figure*}

Training on \BinBF~and \NHj~improved the prediction of \DP~for both TNG and SIMBA on all scales $k = 1.0$--20\,$h$\,Mpc$^{-1}$ as measured by the $\rm{R}^2$ and RMSE/IQR scores (Fig.~\ref{fig:scores}). 
For TNG, we achieved the highest $\rm{R}^2$ score of all experiments at $k = 5\,h$\,Mpc$^{-1}$, with $\rm{R}^2 = 0.923$ (consistent with the lowest RMSE/IQR score). In other words, the RF was able to account for approximately 92\% of the variation in the suppression of the matter power spectrum due to feedback using \BinBF~and \NHj~as training features. This represents $\sim$40\%~improvement over training on \vDBF~at $k = 1.0\,h$\,Mpc$^{-1}$ and $\gtrsim$20\%~improvement over training on \BinBF~and \NHj~at $k = 1.0\,h$\,Mpc$^{-1}$. The trained RF also shows very good performance down to smaller scales, with $\rm{R}^2 = 0.85$--0.9 at $k = 10$--20\,$h$\,Mpc$^{-1}$ when training simultaneously on the baryon fraction and number of halos in different mass bins.
Similar results are obtained for SIMBA, also performing better at $k = 5$--20\,$h$\,Mpc$^{-1}$ compared to larger scales, with $\rm{R}^2$ scores slightly lower than TNG: $\rm{R}^2\approx0.8-0.85$ at $k = 5$--20\,$h$\,Mpc$^{-1}$.
Remarkably, while the vDMS model can only predict \DP~for a given average baryon fraction of massive halos on large scales $k < 1\,h$\,Mpc$^{-1}$, our RF regressor performs better on scales where the impact of feedback on the matter power spectrum becomes the highest. The strongest suppression of power occurs at roughly $k = 10\,h$\,Mpc$^{-1}$ in SIMBA and $k = 20\,h$\,Mpc$^{-1}$ in TNG for their fiducial models, and the RF is able to account for $\sim$80-85\%~of the \DP~variation on these scales.

Fig.~\ref{fig:RF_k5_plots} illustrates in more detail the predicted results versus true values of \DP~when training a RF regressor on different input features at $k = 5\,h$\,Mpc$^{-1}$ for the TNG (top) and SIMBA (bottom) {\bf LH} sets.  The highest $\rm{R}^2$ score was obtained at $k = 5\,h$\,Mpc$^{-1}$ for TNG using \BinBF~and \NHj~as training features, which corresponds to the tighter distribution of blue data points along the one-to-one line of perfect prediction in the top left panel, with reduced scatter compared to the prediction based on high mass halos alone (orange data points).  We find similar trends for SIMBA, with an apparent increase in scatter relative to TNG as expected from the lower $\rm{R}^2$ scores. 
Interestingly, the feature importance analysis (right panels) indicates that the number of halos in the low mass range $M_{\rm halo} = 10^{10.5}$--$10^{11}$\,\Msun~is contributing significantly to improve the \DP~predictions.

\subsection{Interpretation of feature importances}

\begin{figure}
    \centering
    \includegraphics[width=0.99\columnwidth]{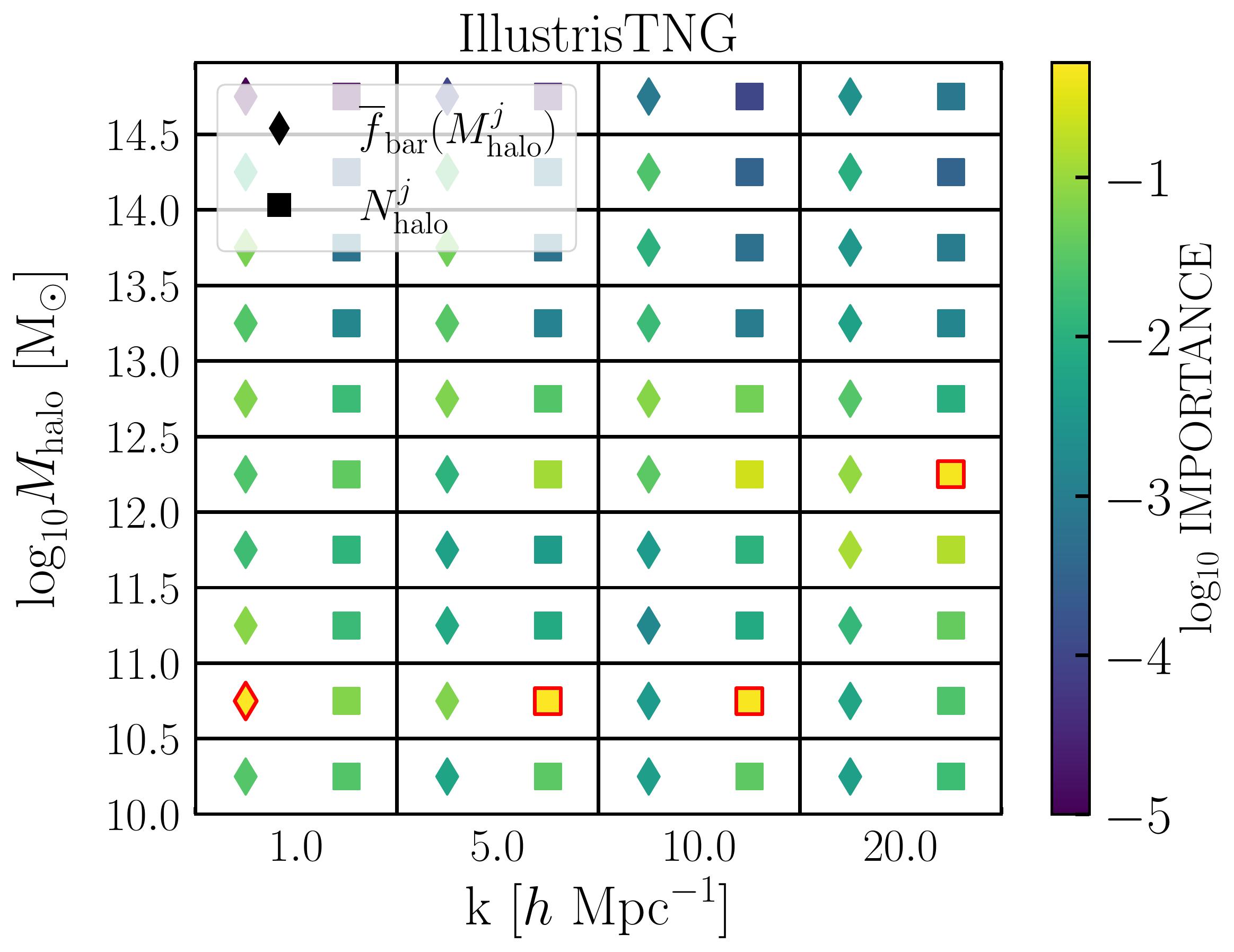}
    \includegraphics[width=0.99\columnwidth]{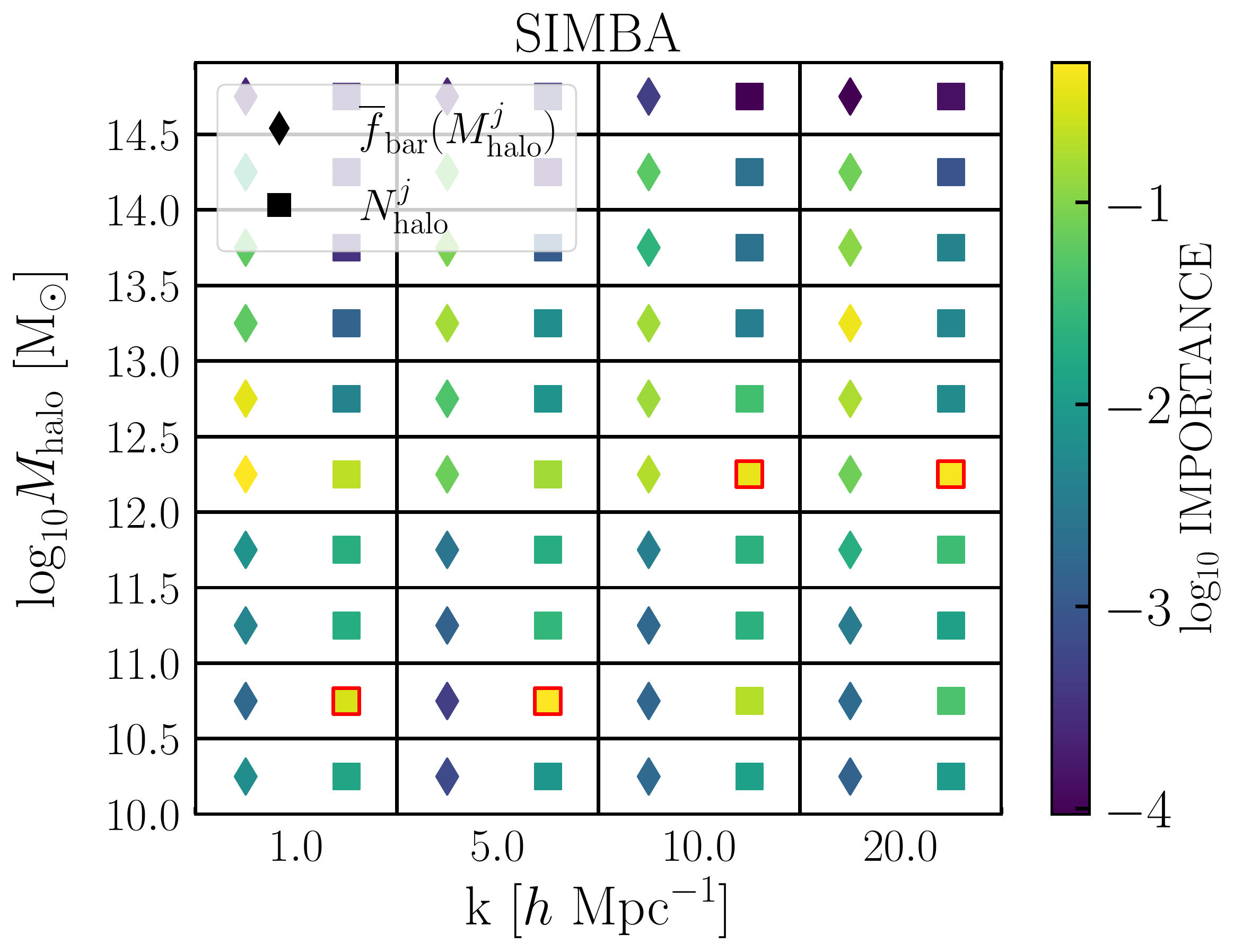}
    \caption{Feature importance summary for RF models trained on \BinBF~and \NH~to predict \DP~at $k=[0.5, 1, 5, 10, 20]\,h$\,Mpc$^{-1}$ for the TNG (top) and SIMBA (bottom) models. For each halo mass bin and $k$-value, the relative importance of the corresponding \BinBF~(diamonds) and \NH~(squares) features is indicated by the color scale. 
    The highest ranked feature for each $k$-value is outlined in red. There is no clear trend of importance in features across scales, suggesting feature importance results are specific to each scale.}
    \label{fig:feature_summary}
\end{figure}

Comparing the feature importances shown in Figs.~\ref{fig:RF_vDMS_vs_binned} and~\ref{fig:RF_k5_plots} for $k = 1.0\,h$\,Mpc$^{-1}$ and $k = 5\,h$\,Mpc$^{-1}$, respectively, it appears that the most informative input features vary with scale. We further investigate the physical properties that inform the prediction of \DP~by the RF in Fig.~\ref{fig:feature_summary}, where we provide a summary of the relative importance of the \BinBF~features (diamonds) and \NHj~features (squares) corresponding to different halo mass bins when predicting \DP~at a given scale $k$. The ranking of features for each $k$-value is indicated by the color scale, which shows the log of the fractional importance assigned by the RF.
The most important ranked feature at each $k$-value is further outlined in red. 
While the RF appears to be learning from the baryon fractions and abundances of halos across the halo mass range, there are some interesting trends that are worth noting. 
The least informative features for both TNG and SIMBA correspond to the most massive halos ($M_{\rm halo} \sim 10^{14}$\,\Msun), which is in contrast to earlier work identifying the baryon fraction in groups and clusters as a primary predictor of \DP~on scales $k < 1\,h$\,Mpc$^{-1}$ (vDMS). Unsurprisingly, the small simulated volumes in CAMELS contain a small number of massive halos, which are thus not optimal as predictors of \DP. In contrast, the RF assigns significant importance to \BinBF~and \NHj~in low- to intermediate-mass halos in the range $M_{\rm halo} \sim 10^{10.5}$--$10^{12.5}$ \Msun~for all $k$-values analyzed.

\begin{figure*}
    \centering
    \includegraphics[scale=0.3]{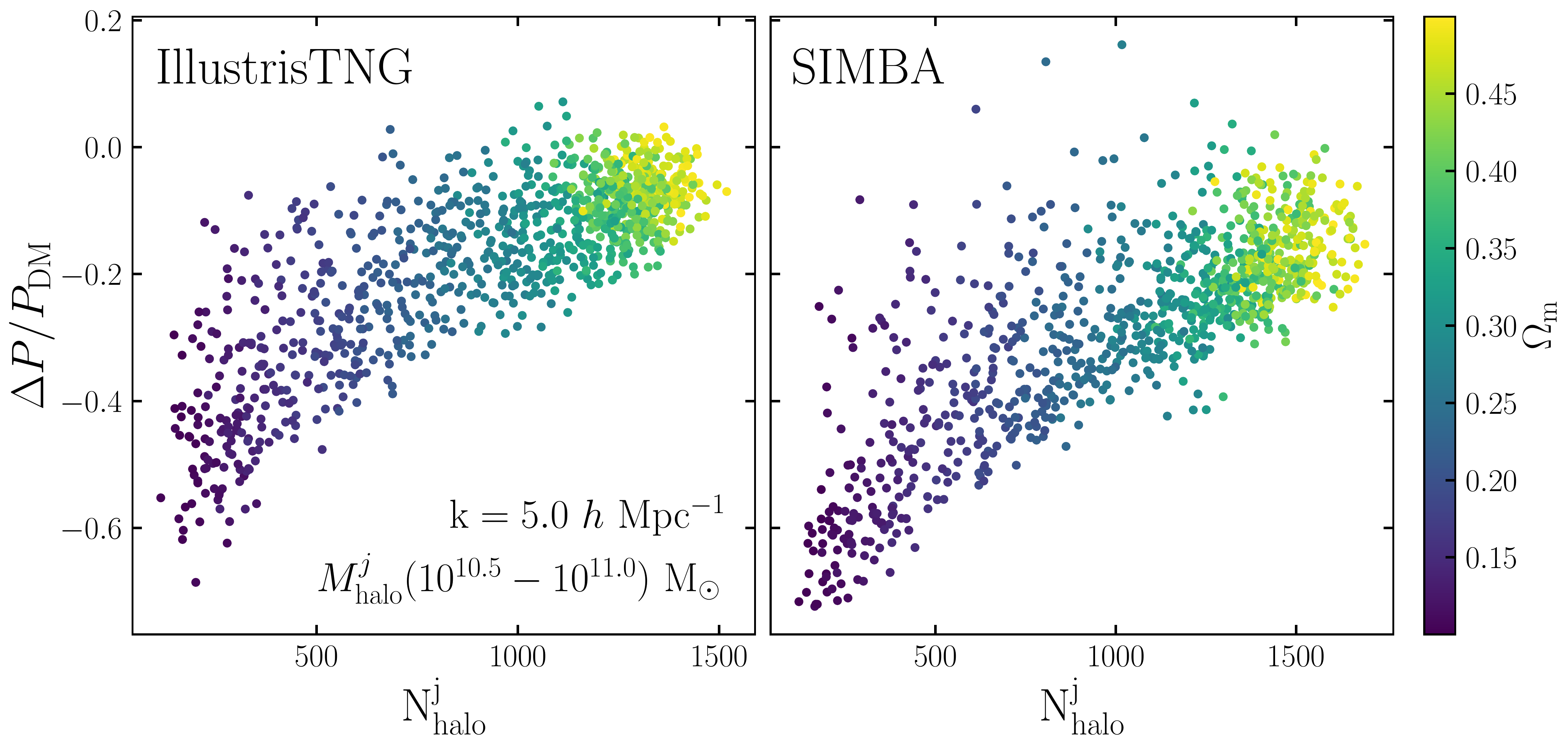}
    \caption{Correlation between power spectrum suppression \DP~and the number \NH~of halos in the mass range $M^{\rm j}_{\mathrm{halo}} = 10^{10.5}$--$10^{11}$ \Msun~at $k=5\,h$\,Mpc$^{-1}$ for TNG (left) and SIMBA (right), color coded by $\Omega_{\rm{m}}$. 
    The number of low mass halos \NH~(highest ranked feature at $k=5\,h$\,Mpc$^{-1}$) is a strong tracer of $\Omega_{\rm{m}}$ in CAMELS  and a good predictor of \DP.
    }
    \label{fig:feature_correlations}
\end{figure*}

Interestingly, the number of halos \NHj~in the mass range $M_{\rm halo} \sim 10^{10.5}$--$10^{11}$\,\Msun~is among the top features identified by the RF across different scales (see also Fig~\ref{fig:RF_k5_plots}).
We explore further the significance of this feature in Fig.~\ref{fig:feature_correlations}, where we show \DP~at $k = 5\,h$\,Mpc$^{-1}$ as a function of \NHj~in this halo mass range for the {\bf LH} sets of TNG (left) and SIMBA (right).  We find that there is a clear correlation between \DP~and the number of low-mass halos for both galaxy formation models, as expected given that \NHj~is identified by the RF as one of the most predictive features.  Physically, a plausible explanation for this correlation is that \NHj~for low mass halos is a strong tracer of $\Omega_{\rm{m}}$, as indicated by the color scale, and $\Omega_{\rm{m}}$ itself is one of the main parameters driving large variations in \DP~at all $k$-values in CAMELS (see Figs.~\ref{fig:phyd_pdm_1param}, \ref{fig:TNG_VDplots}, and~\ref{fig:SIMBA_VDplots}). This is consistent with \cite{2023arXiv230102186P}, which find a simple model is also able to capture information about \DP~given \BF~in low-mass halos and \Om. While increasing the value of $\Omega_{\rm{m}}$ increases systematically the number of halos at all masses \citep[e.g.,][]{camels_presentation}, we note that the correlation between \DP~and \NHj~worsens for higher halo mass bins, which are thus often assigned lower feature importance by the RF.  This can be explained by the number of halos in higher mass bins being more sensitive to cosmic variance and therefore not as good predictors of cosmology for small simulated volumes. On the other hand, the lowest halo mass bin considered here, $M_{\rm{halo}}<10^{10.5}$ \Msun, becomes unresolved in terms of the minimum number of dark matter particles per halo for the higher $\Omega_{\rm{m}}$ values in CAMELS, as depicted in Fig. 8 of \cite{camels_presentation}. The halos in the lowest mass bin in this study have less than 200 dark matter particles. We therefore conclude that \NHj~in the halo mass bin $M_{\rm halo} \sim 10^{10.5}$--$10^{11}$ \Msun~is the least sensitive to cosmic variance and it is a strong predictor of \DP~in part because \NHj~informs the RF about the variation of $\Omega_{\rm{m}}$ in a large number of well resolved halos. 

\begin{figure*}
    \centering
    \includegraphics[scale=0.3]{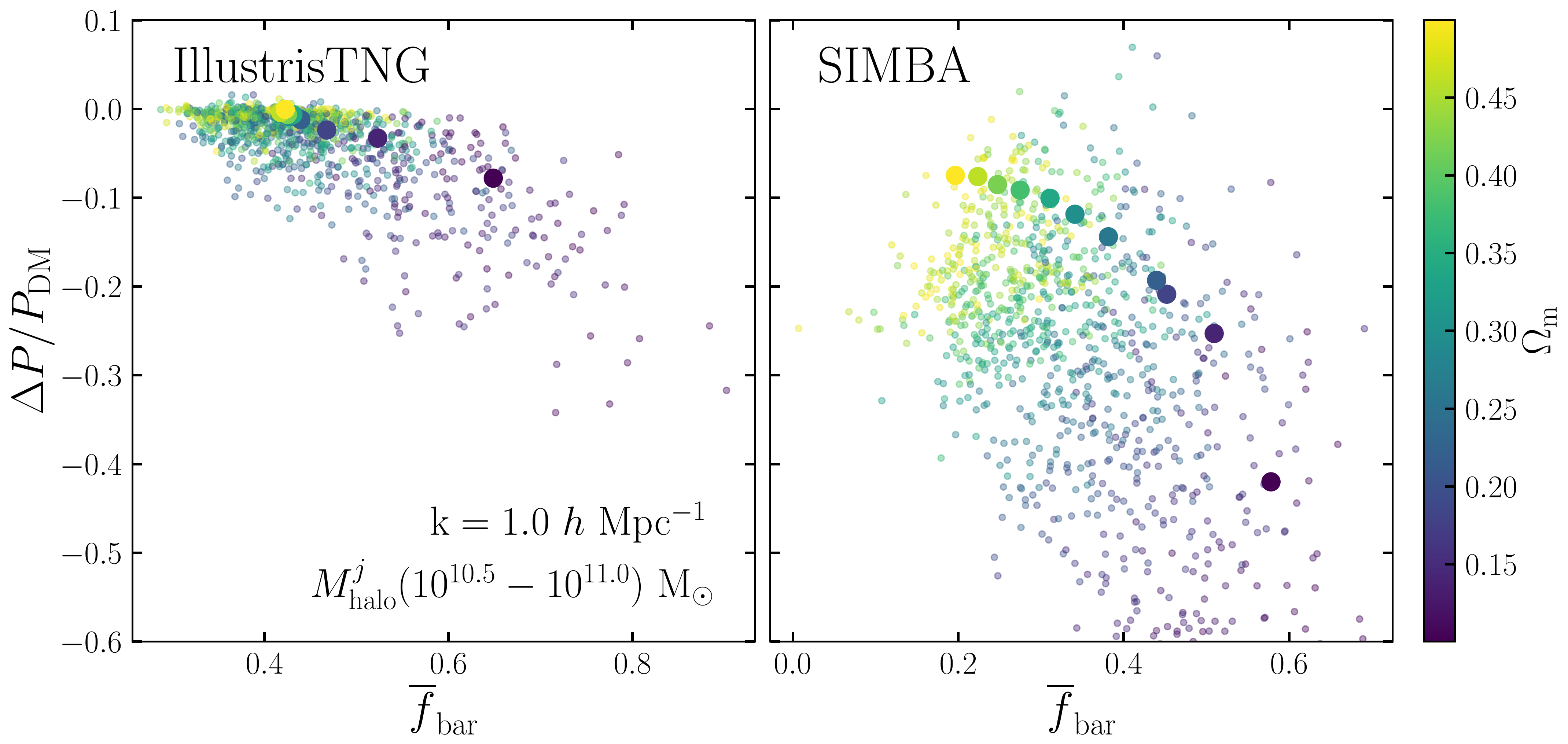}
    \caption{Similar to Figs. \ref{fig:SIMBA_VDplots} and \ref{fig:TNG_VDplots} but here we look only at the dependence of the \DP~-\BF~relation on \Om~for the 
    \BinBF~in the range $M^{\rm j}_{\mathrm{halo}} = 10^{10.5}$--$10^{11}$. Again, the small dots represent the {\bf LH} sets while the large circles are {\bf 1P} sets. We see an inverted trend compared to the vDMS relation which holds for high mass halos, but there is still a strong dependence on \Om.}
    \label{fig:lowmass_vDMS}
\end{figure*}

In order to further examine the significance of tracers of \Om~in low-mass halos as predictors of \DP~we look specifically at the dependance of the \DP~-\BF~relation on \Om~in the halo mass range $M^{\rm j}_{\mathrm{halo}} = 10^{10.5}$ - $10^{11}$\Msun. Results are shown in Fig. \ref{fig:lowmass_vDMS}. Again, the small dots represent the {\bf LH} sets while the large circles are {\bf 1P} sets. We see an inverted trend compared to the vDMS relation which holds for high mass halos and there is still a clear dependence on \Om.
These behaviors (increased suppression with higher \BF~and lower \Om) are consistent with results in the \Om~panels of both Fig. \ref{fig:phyd_pdm_1param}, where we see stronger suppression of the matter power spectrum with higher values of \Om~, and of Fig. \ref{fig:1P_bf_vs_vmassom}, where we see that \BF~increases as \Om~decreases in low-mass halos. We again remind the reader that our computed \BF~ are normalized by $\Omega_{\rm b}$/ \Om. Therefore, lowering the values of \Om~at a fixed $\Omega_{\rm b}$ results in a non-trivial effect on the suppression of power.

\subsection{Marginalizing over galaxy formation physics}
\label{Results:RF_2model}

A key advantage of CAMELS over more standard cosmological simulations performed with a single fiducial galaxy formation model is the ability to train machine learning algorithms to learn fundamental properties of galaxies and the Universe while marginalizing over uncertainties in subgrid physics \cite[e.g.,][]{Villaescusa-Navarro2021_RobustMarginalization,2022JCAP...04..046N,Perez2023_CAMELS-SAM,Shao2022_subhalo_VirialRelation,Shao2023_RobustGNN_halos,Villaescusa-Navarro2021_MultifieldCosmology,Villanueva-Domingo2022}. 
When training a RF on the {\bf LH} simulation set of either TNG or SIMBA to predict \DP~given halo baryon fractions as input features, we are at the same time marginalizing over uncertainties in physical processes represented by the parameter variations introduced in a given galaxy formation model.  

\begin{figure}
    \centering
    \includegraphics[width=0.99\columnwidth]{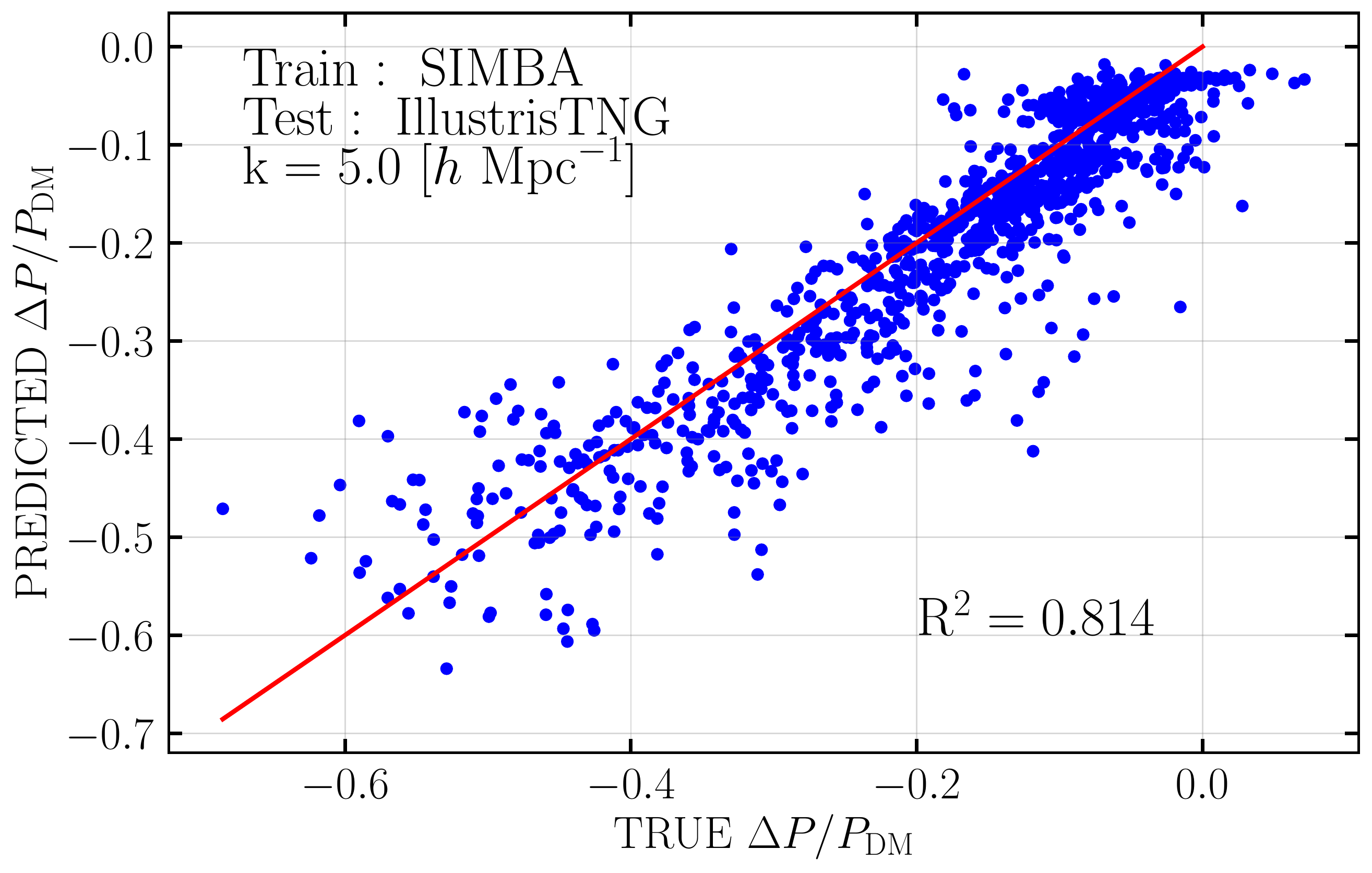}
    \includegraphics[width=0.99\columnwidth]{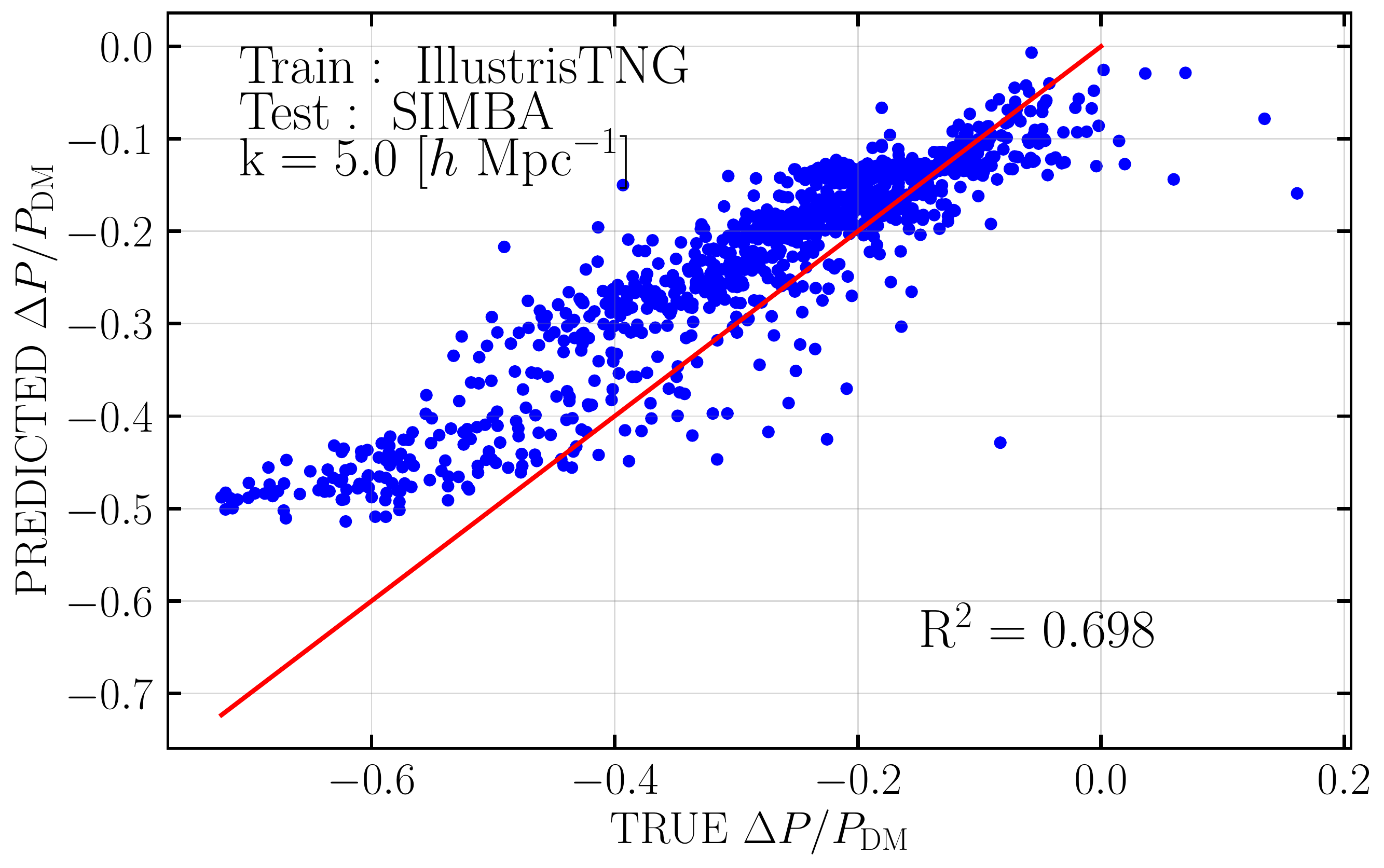}
    \caption{Similar to Fig. \ref{fig:RF_k5_plots} but training a RF on the entire {\bf LH} set of one galaxy formation model and testing on the entire {\bf LH} set of the other for both TNG and SIMBA. We note higher prediction score when training on SIMBA and testing on TNG ($R^2 = 0.814$) compared to training on TNG and predicting on SIMBA ($R^2 = 0.698$), which can be attributed to SIMBA's wider range of \DP~values. The RF tends to under-predict \DP~when trained on SIMBA and over-predict \DP~when trained on TNG due to variations in feedback models which suggests that care must be made if applying this model to data where feedback is not precisely known.
    }
    \label{fig:mix_RF}
\end{figure}

However, evaluating the robustness of the ML model to uncertainties in galaxy formation physics should also consider different implementations and not just variations of parameters within a given subgrid physics implementation. We thus perform a more stringent test of robustness by 
training the RF on the full {\bf LH} set of one galaxy formation model (either SIMBA or TNG) and then testing on the full {\bf LH} set of the other model.  Fig.~\ref{fig:mix_RF} shows the predicted versus true values of \DP~at $k = 5\,h$\,Mpc$^{-1}$ when training a RF using \BinBF~and \NH~from SIMBA and testing on TNG (top) and when training on TNG and testing on SIMBA (bottom).
The top panel of Fig.~\ref{fig:mix_RF} shows that the RF trained on SIMBA can explain $\sim$80\%~of the variation of \DP~at $k = 5\,h$\,Mpc$^{-1}$ when tested on TNG, suggesting that the RF has found a relation between halo baryon fractions and suppression of matter clustering which is relatively robust to galaxy formation model implementation.  On the other hand, the bottom panel of Fig.~\ref{fig:mix_RF} shows that the RF trained on TNG is less robust when tested on SIMBA, and can only explain $\sim$70\% of the variation in \DP~predicted by the SIMBA model.
In this case, we can see that the predicted \DP~is clearly biased high (i.e., less negative) when training on TNG and predicting on SIMBA, implying that the inferred suppression of matter clustering is under-predicted given the halo baryon fractions in SIMBA and the connection to \DP~learned from TNG.  As expected, we see a bias in the opposite direction when training on SIMBA and predicting on TNG (top panel), over-predicting the suppression of power at $k = 5\,h$\,Mpc$^{-1}$.

As seen in Figs.~\ref{fig:SIMBA_VDplots} and~\ref{fig:TNG_VDplots}, the overall range of variation in \DP~is significantly larger in the SIMBA {\bf LH} set compared to the TNG {\bf LH} set. 
Given that the RF cannot predict values outside of the range of the training data, this can explain why the RF trained on TNG is less robust relative to galaxy formation implementation and shows a stronger bias when tested on SIMBA. The biased estimation of \DP~can thus be partially attributed to the RF learning the limits of the range of variation in the training data. Nonetheless, our results suggest that the TNG and SIMBA models may predict different \DP~even when implementing parameters that yield similar \BinBF, implying a non-unique relation between halo baryon fractions and impact on matter clustering.

\subsection{Supplementary studies}

In order to further investigate the performance of training the RF on halo abundances, we performed several supplementary studies comparing our fiducial models to new models which include cosmological parameters as training features. The results of these experiments are presented in Appendix~\ref{appendix}, where we conclude that our fiducial model, utilizing \BinBF~and \NHj~across the full range of halo masses, extracts important cosmological information as it is in close agreement (within $1\%$) with a model trained on \BinBF~and \Om. However, in the case of high mass halos, halo abundance is not a sufficient proxy for cosmological information. We also found that limiting the study to intermediate mass halos resulted in comparable results to our fiducial model when the RF was also directly provided \Om~as a training feature in addition to the \BF~of intermediate mass halos.

\begin{figure*}
\centering
    \includegraphics[width=0.8\textwidth]{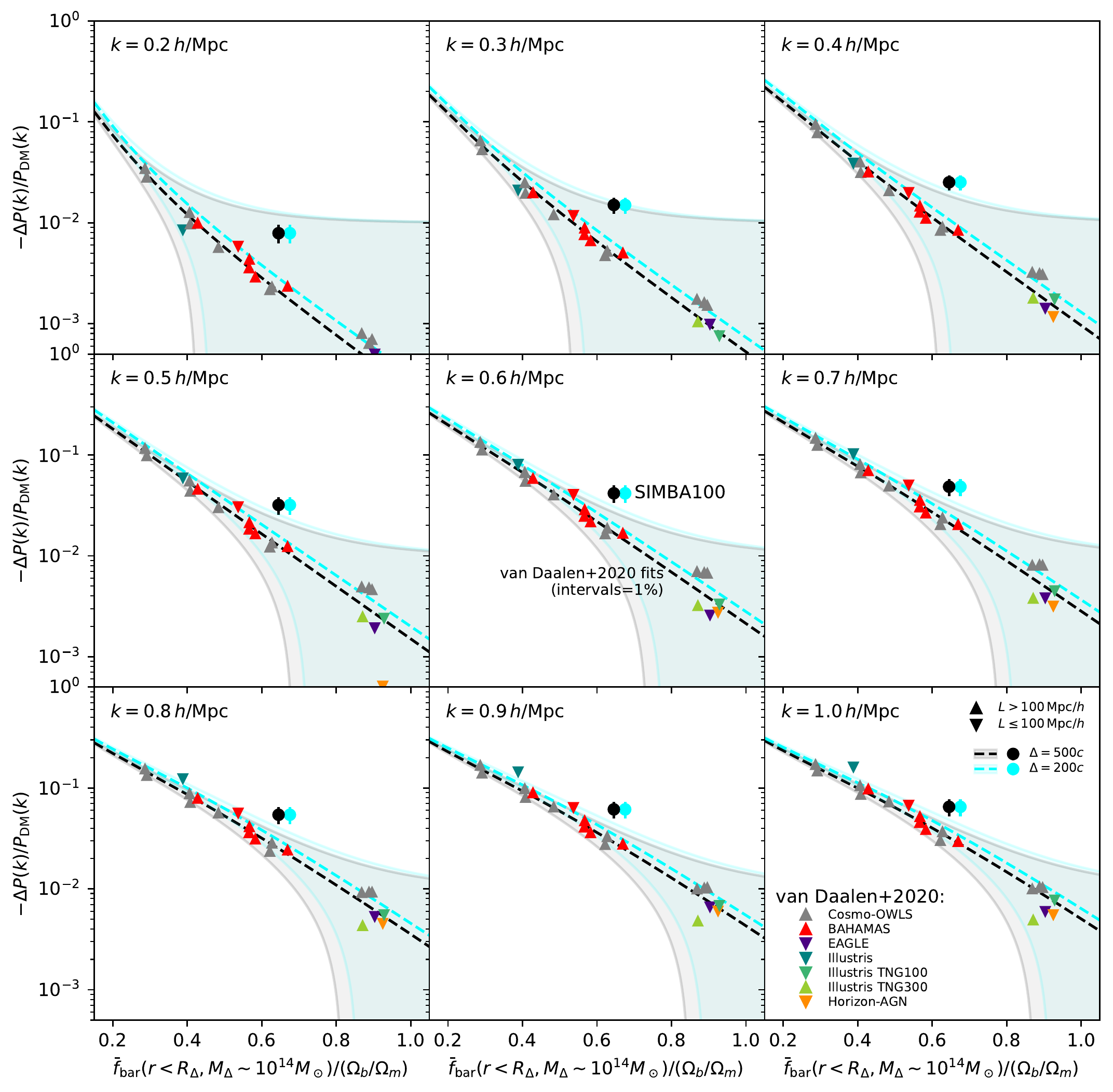}
    \caption{Comparison of the power spectrum suppression in the original SIMBA simulation with the results from vDMS. The SIMBA data points are circles, while the measurements from vDMS are reproduced as triangles. We also show the fitting functions from vDMS as dashed lines as well as 1\% range of variations. 200c and 500c mass definitions are shown in different colors. The vertical error bars have been estimated by splitting the $100\,h^{-1}\text{Mpc}$ SIMBA simulation into 8 sub-volumes. Statistical error bars on the horizontal axis are small ($\sim 10^{-3}$ as estimated using jack-knife), but there are somewhat larger systematic errors from the halo finding ($\sim 10^{-2}$, as estimated by running the analysis with FOF and Rockstar). SIMBA appears to deviate from the best-fit relation of vDMS.
    }
\label{fig:vDSIMBA}
\end{figure*}

Based on a suite of simulations of substantially larger volumes, vDMS identified a tight relationship
between mean baryon fraction in massive halos and baryonic power spectrum suppression at $k < 1\,h$\,Mpc$^{-1}$. While some degree of correlation between these quantities is expected, as we find here, the extremely small scatter in the
observed relationship was remarkable given the variety of galaxy formation models compared by vDMS.
Since we observed some hints at a deviation
from the vDMS relationship in CAMELS, particularly for the parameter variations based on the SIMBA model, we perform a direct comparison of the original $100\,h^{-1}\text{Mpc}$ SIMBA volume to the vDMS relation in Fig.~\ref{fig:vDSIMBA}. 
Here, we show the new SIMBA data points as circles, while the measurements considered in vDMS are shown as triangles.
We have verified our pipeline by running it on IllustrisTNG-300, achieving excellent agreement with
the vDMS measurements.
As can be seen, SIMBA constitutes a considerable outlier and does not fall within the 1\%~interval
around the vDMS fit. The only other simulation scattering that far is Illustris which, however, does
not reproduce the observed baryon fraction.
It is known that the feedback prescription in SIMBA is unique in its ability to re-distribute baryons
across large scales \citep{Borrow2020,2023arXiv230711832G}, which could explain the observed deviation from the vDMS
relation. However, SIMBA generally does not do worse in reproducing observational relationships than
the other major hydrodynamic simulations. Thus, these results suggest the possibility that the vDMS relation only
holds in a sub-space of simulations while there exists at least one dimension along which deviations occur.

%%%%%%%%%%%%%%%%%%%%%%%%%%%%%%%%%%%%%%%%%%%%%%%%%%
\section{Summary and discussion}
\label{summary}

In this paper, we have investigated how baryonic physics affects the clustering of matter relative to N-body simulations and its relation to the baryonic content of halos using thousands of cosmological hydrodynamic simulations from the CAMELS project \citep{camels_presentation}.
In the first part of the paper, we examined how variations of individual cosmological parameters ($\Omega_{\rm m}$ and $\sigma_8$) and feedback parameters (controlling the efficiency of large-scale outflows driven by SNe and AGN) impact the total matter power spectrum, the mean halo baryon fraction as a function of halo mass, and, motivated by \cite{2020MNRAS.491.2424V} (vDMS), the connection between the suppression of clustering, \DP, and the mean baryon fraction of massive halos.

The small simulated volumes in CAMELS complicate a direct comparison between our results and vDMS, where in addition to cosmic variance affecting \DP~we also lack halos massive enough to evaluate the mean baryon fraction under the same conditions. Therefore, in the second part of the paper, we have presented a set of machine learning experiments as an extension to what was done in vDMS, training a random forest (RF) regressor on features including the mean baryon fraction and introducing the abundance of halos across the mass range $10^{10} \leq M_\mathrm{halo}/{\rm M}_{\odot}\,h^{-1} < 10^{15}$ to predict \DP~from linear to highly non-linear scales ($k = 1.0$--20.0\,$h\,\mathrm{Mpc}^{-1}$). By utilizing halo abundance as a training feature, the RF learns about the cosmic variance present in the simulations without explicitly knowing the underlying cosmology.

Throughout the paper, we have made use of the CAMELS simulation suites performed with the TNG \citep{2018MNRAS.475..648P, 2018MNRAS.475..676S, 2018MNRAS.475..624N, 2018MNRAS.477.1206N, 2018MNRAS.480.5113M} and SIMBA \citep{2019MNRAS.486.2827D} models to understand the dependence of results and the robustness of the trained machine learning models to changes in the specific galaxy formation physics implementation.

Our main findings can be summarized as follows: 
 
\begin{itemize} 

\item In agreement with previous work \citep[e.g.,][]{2011MNRAS.415.3649V,2018MNRAS.480.3962C,2019OJAp....2E...4C,camels_presentation}, we find that baryonic physics can profoundly affect the total matter power spectrum all the way to scales $k < 0.5\,h\,\mathrm{Mpc}^{-1}$, and the magnitude of this effect is highly dependent on the details of the galaxy formation implementation and variations of cosmological and astrophysical parameters. \\ 

\item The suppression of power, |\DP|, increases systematically with decreasing $\Omega_{\rm m}$ at fixed $\Omega_{\rm b}$, with baryons contributing a higher fraction of the total matter content and feedback more efficiently spreading matter over larger scales relative to N-body simulations. Varying $\sigma_8$ at fixed galaxy formation physics does not drive systematic variations in \DP~when measured on the small $(25\,h^{-1}\mathrm{Mpc})^3$ volumes simulated in CAMELS. \\ 

\item  Increasing AGN feedback efficiency generally drives higher suppression of matter clustering, in agreement with previous work \citep[e.g.; vDMS;][]{2022JCAP...04..046N}, with the strongest effect seen for high-speed jets in SIMBA which are able to spread a substantial amount of baryons over scales of several Mpc \citep{Borrow2020,2023arXiv230711832G}. The qualitative effect of stellar feedback on matter clustering is more dependent on galaxy formation model, which can either suppress or enhance power on different scales depending on the interplay between stellar and AGN feedback. Stronger stellar feedback often results in weaker overall suppression of matter clustering by suppressing black hole growth and therefore the effective efficiency of AGN feedback \citep{2011MNRAS.415.3649V,2022JCAP...04..046N,2023arXiv230711832G}.\\

\item Halo baryon fractions \BF~are very sensitive to galaxy formation model, with TNG producing systematically more baryon-rich halos compared to SIMBA for a broad range of parameter variations. Higher AGN feedback efficiency generally decreases halo baryon fractions, but the extent of the effect and the affected halo mass range depend on model details.  Increasing the strength of stellar feedback can either decrease or increase the baryon fraction depending on the non-linear coupling of stellar feedback and black hole growth. Halo baryon fractions are also very sensitive to changes in cosmology. Increasing $\Omega_{\rm m}$ (at fixed $\Omega_{\rm{b}}$) or $\sigma_8$ systematically decreases the baryon fraction of halos (normalized by $\Omega_{\rm b}/\Omega_{\rm m}$), indicating a non-trivial response of feedback to changes in the amount of baryons relative to dark matter and the growth history of halos. \\

\item  We find a broad correlation between the amount of suppression of the matter power spectrum \DP~and the baryon fraction of massive halos \BF, indicating that the feedback mechanisms responsible for evacuating gas from massive halos also dominate the impact of baryonic effects on matter clustering. These results are in broad agreement with vDMS, but the thousands of simulations in CAMELS produce significantly larger scatter in the \DP--\BF~relation.  Cosmic variance alone can significantly affect the matter power spectrum on our $(25\,h^{-1}\mathrm{Mpc})^3$ simulated volumes, but the complex trends seen for the impact of individual cosmological and feedback parameter variations on \DP~and \BF~suggest that the vDMS model predicting \DP~given only \BF~for massive halos is not general enough to include every plausible feedback model. \\

\item  Predicting the impact on matter clustering based only on the mean baryon fraction of massive halos using the vDMS \DP--\BF~relation is not possible given the broad range of galaxy formation models and the impact of cosmic variance in CAMELS.  However, we have demonstrated that a RF regressor trained on CAMELS is able to extract information from halos across the full mass range $10^{10} \leq M_\mathrm{halo}/{\rm M}_{\odot} < 10^{15}$ to estimate the suppression of the matter power spectrum on scales $k = 1.0$--20\,$h\,\mathrm{Mpc}^{-1}$. We are thus not only extracting information from low-mass halos but also predicting \DP~in the highly non-linear regime, significantly extending the range of scales $k < 1\,h$\,Mpc$^{-1}$ where the vDMS model can be applied. \\

\item Using the mean halo baryon fraction and abundance in different halo mass bins as input features, the RF regressor was able to account for $\sim$80--85\%~of the \DP~variation occurring on scales $k = 10$--20$\,h$\,Mpc$^{-1}$ where the impact of feedback on the matter power spectrum becomes the highest. At 
$k=5.0\,h$Mpc$^{-1}$, our best model was able to explain
$\sim$92\%~of the variance in the suppression of power due to feedback when training on the TNG model. However, the same model can only explain $\sim$70\% of the variation in \DP~and tends to under-predict the suppression of matter clustering when tested on the SIMBA simulations, indicating that the RF is only moderately robust  relative to changes in the underlying galaxy formation implementation. Training on SIMBA increases the robustness of the model owing to its larger range of variation in \DP~compared to TNG, but in this case the RF tends to over-predict |\DP| when tested on TNG.
These results suggest that the lack of a universal relation between halo baryon fractions and impact on matter clustering and emphasize the need to construct models that are robust against assumptions in baryonic physics \citep[e.g.,][]{Villaescusa-Navarro2021_RobustMarginalization,2022JCAP...04..046N,Shao2023_RobustGNN_halos}. \\

\item The original SIMBA volume constitutes a considerable outlier to the vDMS relation and does not fall within the 1\%~interval
around the vDMS fit.
\\

\item Our fiducial model, utilizing \BinBF~and \NHj~across the full range of halo masses, extracts important cosmological information as it is in close agreement (within $1\%$) with a model trained on \BinBF~and \Om~(Appendix \ref{appendix}).\\

\end{itemize}

A unique advantage of CAMELS relative to previous work is that it performs simulations for different baryonic physics implementations and a broad range of cosmological and feedback parameter variations, providing a data-set sufficiently large to train machine learning algorithms for a variety of applications \citep{camels_presentation}.
However, an important limitation of CAMELS is the small volume of each simulation realization, $L_\mathrm{box}=25\,h^{-1}\mathrm{Mpc}$, with important implications for this work.  
Given the small box sizes, the matter power spectrum is sensitive to the specific initial conditions in each realization, and the impact of baryonic effects further depends on stochastic processes related to feedback operating on a limited number of massive halos.  As a result, cosmic variance represents a challenge to infer the suppression of matter clustering \DP~given only the baryon fraction of massive halos. Previous works in CAMELS have devised strategies to correct for the noise introduced by cosmic variance. 
When training a neural network on electron density auto-power spectra to predict $\Omega_{\rm m}$, \citet{2022JCAP...04..046N} constructed a cosmic variance parameter based on the distribution of halo masses in each realization, improving the predictions significantly when introduced as an additional training feature. \citet{Thiele2022} used spectral distortion measurements to constrain baryonic feedback and applied a correction factor to the Compton-y distortion by comparing expected values from a simple halo model evaluated for the halo mass function in each CAMELS simulation compared to that of a standard halo mass function. In our RF experiments, introducing the number of halos in each mass bin, \NH~or \NHj, as input features (i.e., basically the halo mass function) improves the accuracy of the predictions significantly.  The number of low-mass halos was one of the most predictive features identified by the RF, which can be understood as a strong tracer of $\Omega_{\rm m}$ (Fig. \ref{fig:feature_correlations}). Halo abundance further serves as a feature which 
provides the RF algorithm information about cosmic variance,
in agreement to previous works. Furthermore, cosmic variance in CAMELS is lower in low-mass halos than high-mass halos, which may account for the predictive power of \NHj in low-mass halos.

One final result that we found puzzling, was the inversion of the \DP~- \BF~ relation and its dependence on \Om~in low mass halos. It is unclear why a lower value in \Om~ at a fixed $\Omega_{\rm b}$ would result in higher \BF~and increased suppression in the total matter spectrum in low-mass halos. The dependence on \DP~on baryonic feedback in low-mass halos will need to be further investigated in order to thoroughly explain these results. It will also be interesting to perform the RF training on a fixed, or more narrow range of \Om~to study how well the RF can really learn the different effects of baryonic feedback, but we leave this to a future study.

\section{Acknowledgments}
We wish to thank the anonymous referee for insightful comments which aided in improving this work. We also thank Dylan Nelson who likewise provided valuable comments.
DAA acknowledges support by NSF grants AST-2009687 and AST-2108944, CXO grant TM2-23006X, Simons Foundation Award CCA-1018464, and Cottrell Scholar Award CS-CSA-2023-028 by the Research Corporation for Science Advancement.

\section{Data Availability}
The simulations used in this work are part of the CAMELS public data release \citep{2023ApJS..265...54V} and are available at  \url{https://camels.readthedocs.io/en/latest/}.

%%%%%%%%%%%%%%%%%%%%%%%%%%%%%%%%%%%%%%%%
%%%%%%%%%%%%%%%%%%%%%%%%%%%%%%%%%%%%%%%%
%clearpage

\bibliographystyle{mnras}
\bibliography{CAMELS}
%\printbibliography

\appendix
\section{Results from supplementary studies}
\label{appendix}
\begin{figure*}
    \centering
    \includegraphics[scale=0.4]{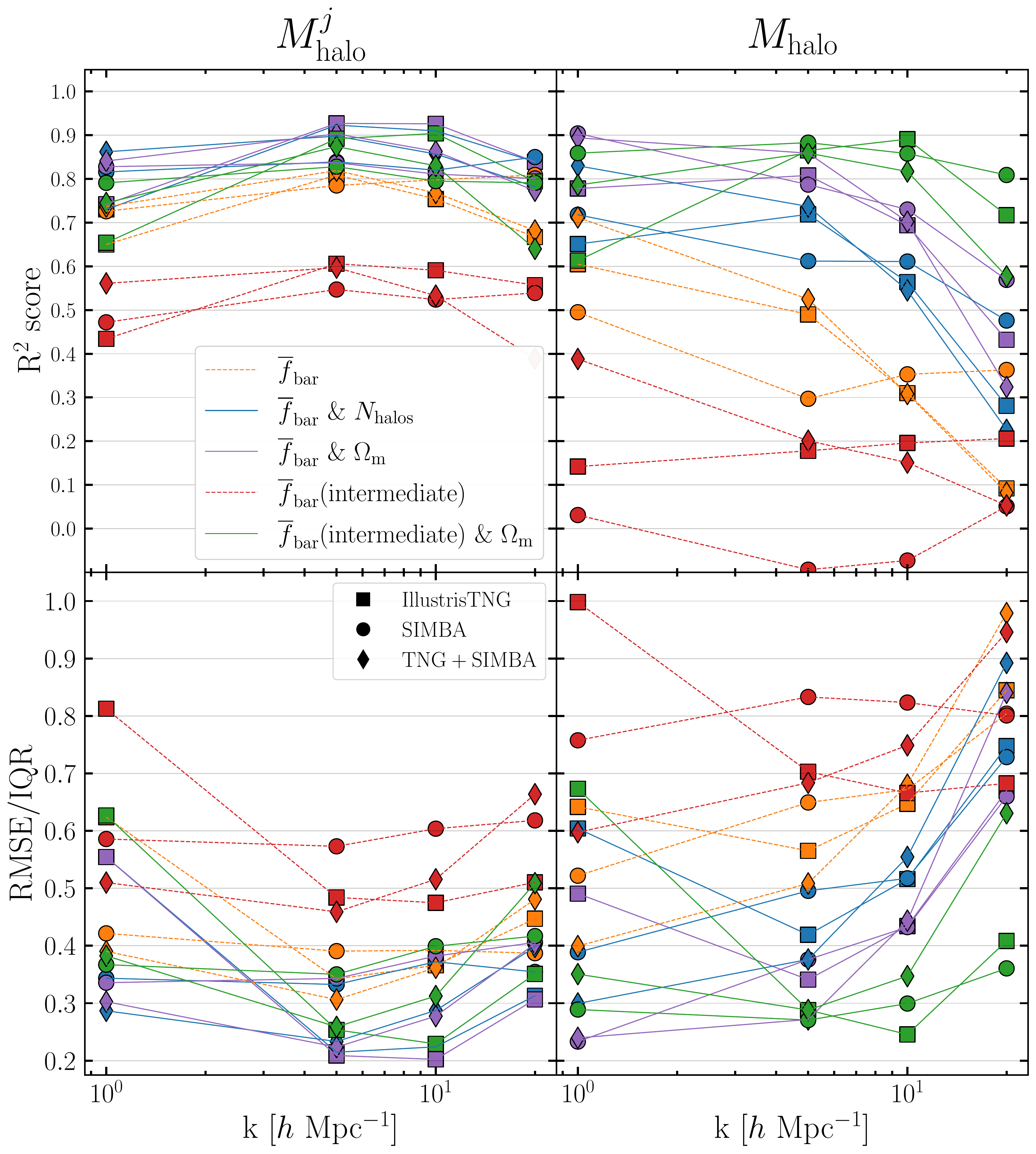}
    \caption{Performance scores for additional RF experiments. Results from the main studies in this paper are still shown in blue and orange as in Fig. \ref{fig:scores}. Models are trained on mean baryon fraction with some models having additional training features of halo abundance or \Om~as indicated in the legend. \textbf{\textit{Left:}} Results shown in orange, blue and purple are for models trained on \BinBF~of the full range of halo masses, while those shown in red and green are for \BinBF~of intermediate mass halos, $M^j_{\rm halo}\in [10^{12}-10^{14})$. \textbf{\textit{Right:}} Results shown in orange, blue and purple are for models trained on \vDBF. Results shown in red and green are for those trained on \BF~of $M_{\rm halo}\in [10^{12}-10^{14})$.}
    \label{fig:scores_supplamental}
\end{figure*}

In order to further investigate the performance of training the RF on halo abundances, we performed several supplementary studies comparing our fiducial models to new models which include cosmological parameters as training features. The results for these can be seen in Fig. \ref{fig:scores_supplamental}. We continue to show the main results from this work in orange and blue, as done in Fig. \ref{fig:scores}, in order to provide a more clear comparison to the supplementary studies. We find the following:
\begin{itemize}
    \item We introduced the training feature \Om~as a substitute to halo abundance (\NH~or \NHj). Scores for these results are shown in purple in Fig. \ref{fig:scores_supplamental}. In the left panels, scores for training on \BinBF~and \Om~(purple) and those for our fiducial model (\BinBF~and \NHj, blue) are within $\approx 1\%$ difference. This indicates that the RF was able to extract important cosmological information given the full range of halo masses and halo abundances. However in the right panel, we see that in the RF model trained on \vDBF~ and \Om~ resulted in $\approx 20\%$ on average improvement in scores compared to that trained on \vDBF~and \NH. This indicates that in the case of high mass halos in CAMELS, halo abundance is not a sufficient proxy for cosmological information.
    \item We introduced $\sigma_8$ as a third training feature (i.e. models trained on mean baryon content, \Om and $\sigma_8$). There were no improvements to the results from using $\sigma_8$ as an additional training feature and results are not shown in Fig.\ref{fig:scores_supplamental} in order to maintain visual clarity.
    \item We trained the RF on mean baryon content and $\sigma_8$. The performance was poor compared to the models trained on mean baryon content and \Om~and results are not shown in order to maintain visual clarity.
\end{itemize}

We further explored the relation between \DP~and mean baryon content in a regime of more intermediate mass halos. Because we are limited in the number of realizations with halos of mass $M_{\rm halo} \geq 10^{13.5}$\Msun, we repeated the additional RF studies described above, this time with \BF~of $M_{\rm halo}\in [10^{12}-10^{14})$\Msun~as a training feature. We also created 8 bins of $M^j_{\rm halo}\in [10^{12}-10^{14})$. The results from training the RF on baryon content from intermediate mass halos can be compared to those using the full range of halo masses. The scores from these results are shown in Fig. \ref{fig:scores_supplamental}.
\begin{itemize}
    \item We found that training the RF on the mean baryon content of intermediate mass halos alone (red) resulted in significantly poorer scores than training on either \vDBF~(right panels shown in orange) or on the \BinBF~of the full range of halo masses (left panels shown in orange).

    \item The \BF~of intermediate mass halos and \Om~model (right panels shown in green) resulted in significantly improved performance of the RF ability to predict \DP~compared to the \vDBF~and \NH~model (right panels shown in blue). Improved scores average $\approx 33\%$ difference. The results for this improved model are comparable to those shown in green in the left panels, which are for a model using \Om~and \BinBF~for the range of intermediate masses. This suggests that there is no meaningful information to be extracted by binning intermediate halo masses to obtain the mean baryon content per mass bin vs the \BF~averaged over all intermediate mass halos. Furthermore, our fiducial model (RF trained on \BinBF~and \NHj~from the full range of halo masses, shown in blue in the left panels) is only, on average, $6\%$ better than that trained on \BF~of intermediate mass halos and \Om, suggesting the latter is promising and worth further investigation as it is a simpler model.
    \item We further limited the halo mass range of intermediate mass halos to $M_{\rm halo}\in [10^{12}-10^{13})$\Msun~and repeated our experiments. Scores were on average $\approx 10\%$ worse than those discussed in this section and results are not shown in order to maintain visual clarity. However, this further demonstrates that there is important information to be extracted from the baryon content of more massive halos.
\end{itemize}

\label{lastpage}
\end{document}